\documentclass[preprint]{aastex61}

\usepackage{amsmath}
\usepackage{graphicx}
\usepackage{natbib}
\usepackage{ulem}
\usepackage{bm}

\usepackage{xspace}
\usepackage{hyperref}

\usepackage{mathtools}

\usepackage{multirow}

\newtheorem{IVP}{IVP}
\newtheorem{EVP}{EVP}
\newtheorem{GP}{GP}
\newtheorem{SP}{SP}

\defcitealias{2022ApJ...928...33L}{Paper~I}
\defcitealias{2007PhPl...14e2101A}{AG07}
\defcitealias{1996ApJ...472..398B}{BDBG96}
\defcitealias{1989CourantHilbert}{CH89}
\defcitealias{1983SoPh...88..179E}{ER83}
\defcitealias{2015ApJ...801...23L}{LN15b}
\defcitealias{2015ApJ...810...87L}{LN15a}
\defcitealias{2014ApJ...789...48O}{ORT14}
\defcitealias{2015ApJ...806...56O}{ORT15}

\newcommand\bigO{\ensuremath{\mathcal{O}}}

\renewcommand{\vec}[1]{\ensuremath{\bm{#1}}}

\newcommand{\Exp}[1]{\ensuremath{{\rm e}^{#1}}}

\newcommand{\innerpClosed}[2]{\ensuremath{\left<#1 |#2\right>^{(d)}}}
\newcommand{\innerpOpen}[2]{\ensuremath{\left<#1 |#2\right>^{(\infty)}}}

\newcommand{\Ac}{\ensuremath{A_{\rm c}}}
\newcommand{\As}{\ensuremath{A_{\rm s}}}
\newcommand{\Alf}{Alfv$\acute{\rm e}$n}

\newcommand{\Schrod}{Schr{\"o}dinger}

\newcommand{\kappae}{\ensuremath{\kappa_{\rm e}}}
\newcommand{\ki}{\ensuremath{k_{\rm i}}}
\newcommand{\ke}{\ensuremath{k_{\rm e}}}

\newcommand{\mui}{\ensuremath{\mu_{\rm i}}}
\newcommand{\mue}{\ensuremath{\mu_{\rm e}}}

\newcommand{\omgl}{\ensuremath{\omega_l}}
\newcommand{\omgcrit}{\ensuremath{\omega_{\rm crit}}}

\newcommand{\va}{\ensuremath{v_{\rm A}}}
\newcommand{\vai}{\ensuremath{v_{\rm Ai}}}
\newcommand{\vae}{\ensuremath{v_{\rm Ae}}}
\newcommand{\vhatd}{\ensuremath{\hat{v}^{(d)}}}

\newcommand{\rhoi}{\ensuremath{\rho_{\rm i}}}
\newcommand{\rhoe}{\ensuremath{\rho_{\rm e}}}

\submitjournal{ApJ}

\shorttitle{Sausage Perturbations in Coronal Slabs}
\shortauthors{Wang et al.}
\begin{document}

\title{Standing Sausage Perturbations in Solar Coronal Slabs with 
{Continuous Transverse Density Profiles}: 
cutoff wavenumbers, evanescent eigenmodes, and oscillatory continuum}

\correspondingauthor{Bo Li}
\email{bbl@sdu.edu.cn}

\author{Zexing Wang}
\affiliation{Shandong Provincial Key Laboratory of Optical Astronomy and Solar-Terrestrial Environment,
   Institute of Space Sciences, Shandong University, Weihai 264209, China}

\author{Bo Li}
\affiliation{Shandong Provincial Key Laboratory of Optical Astronomy and Solar-Terrestrial Environment,
   Institute of Space Sciences, Shandong University, Weihai 264209, China}

\author{Shao-Xia Chen}
\affiliation{Shandong Provincial Key Laboratory of Optical Astronomy and Solar-Terrestrial Environment,
   Institute of Space Sciences, Shandong University, Weihai 264209, China}

\author{Mijie Shi}
\affiliation{Shandong Provincial Key Laboratory of Optical Astronomy and Solar-Terrestrial Environment,
   Institute of Space Sciences, Shandong University, Weihai 264209, China}

\begin{abstract}
{The lack of observed sausage perturbations in solar 
    active region loops is customarily attributed to}
    the relevance of cutoff axial wavenumbers and the consequent absence
    of trapped modes (called ``evanescent eigenmodes'' here).
However, { some recent eigenvalue problem studies} 
    yield that cutoff wavenumbers may disappear for { those equilibria}
    where the external density varies sufficiently slowly, thereby casting
    doubt on { the rarity of candidate} sausage perturbations. 
We examine the responses of straight, transversely structured, coronal slabs
    to small-amplitude sausage-type perturbations that excite axial fundamentals
    by solving
    the pertinent initial value problem with eigensolutions
    for a closed domain. 
The density variation in the slab exterior is dictated by
    some steepness parameter $\mu$, 
    and cutoff wavenumbers are theoretically expected to be present (absent)
    when $\mu \ge 2$ ($\mu < 2$).
However, our numerical results show no qualitative difference
    in the system evolution when $\mu$ varies, despite the differences in
    the modal behavior.
Only oscillatory eigenmodes are permitted when $\mu \ge 2$.
{ Our discrete eigenspectrum becomes increasingly closely spaced
    when the domain broadens, and an oscillatory continuum
    results for a truly open system}.
Oscillatory eigenmodes remain allowed and dominate the system evolution 
    when $\mu <2$.
{ We show that the irrelevance of cutoff wavenumbers does not mean that
	all fast waves are evanescent.
Rather, it means} that an increasing number of evanescent eigenmodes
    emerge when the domain size increases. 
We conclude that sausage perturbations remain difficult to detect even
    for the waveguide formulated here. 
\end{abstract}

\keywords{magnetohydrodynamics (MHD) --- Sun: corona --- Sun: magnetic fields  --- waves}

\section{INTRODUCTION}
\label{sec_intro}

Solar coronal seismology (SCS) is an enterprise that unites theories
    and observations, as was clear
    in the early studies where SCS was practiced implicitly
    \citep[e.g.,][]{1959ApJ...130..215B,1969ApJ...155L.117P,1970A&A.....9..159R}
    or advocated explicitly 
    \citep[e.g.,][]{1970PASJ...22..341U,1984ApJ...279..857R}.
Modern SCS remains to rely on the wealth of measurements of
    low-frequency waves and oscillations in the structured solar corona
    (see the reviews by e.g.,
    \citealt{2005LRSP....2....3N,2012RSPTA.370.3193D,2020ARA&A..58..441N}).  
{ Also key to SCS} is the increasing
    refinement of magnetohydrodynamic (MHD)
    theories for waves in structured media
    (see the reviews by e.g., \citealt{2000SoPh..193..139R,2011SSRv..158..289G,2013SSRv..175....1M,2016SSRv..200...75N}; also the textbooks
    by \citealt{2019CUP_goedbloed_keppens_poedts,2019CUP_Roberts}).       
{ Such terms as kink and sausage modes have been
    in routine use} since their introduction to the solar context for
    cylindrical (\citealt{1983SoPh...88..179E}; see also
    \citealt{1975IGAFS..37....3Z,1982SoPh...75....3S,1986SoPh..103..277C})
    and slab equilibria 
    (\citealt{1982SoPh...76..239E}; see also \citealt{1978ApJ...226..650I,1979ApJ...227..319W}).
Observationally, kink waves and oscillations have been frequently imaged in
    extreme ultraviolet (EUV) in both magnetically open
    \citep[see the review by][and references therein]{2021SSRv..217...76B}
    and magnetically closed structures
    (see \citealt{2021SSRv..217...73N} for a recent review;
    see also \citealt{2015A&A...577A...4Z,2019ApJS..241...31N} for catalogs).
{ However, the reported instances of coronal fast sausage waves
    are only sporadic and exclusively connected to quasi-periodic pulsations (QPPs) in 
    solar flares} 
    \citep[see e.g.,][for recent reviews]{2016SoPh..291.3143V,2018SSRv..214...45M,2020SSRv..216..136L,2021SSRv..217...66Z}.
In particular, they have not been reported or even implicated in oscillating 
    active region (AR) loops to our knowledge. 
This study is intended to address { the rarity of fast sausage waves
    in AR loops, 
    emphasizing conceptual understandings} rather than digging
    into the seismological potential.

It proves necessary to detail the pertinent nomenclature. 
We adopt linear, pressureless, ideal, MHD throughout, 
   { focusing on} straight, field-aligned equilibria
   that are structured only { in one transverse direction}. 
{ Only axial standing sausage waves are of interest, with the axial
   wavenumber denoted by $k$}. 
The necessary terms are worded in a way that they apply to
   both slab and cylindrical configurations, in which some 
   interior (denoted by the subscript~${\rm i}$) is density-enhanced relative
   to its exterior (subscript~${\rm e}$). 
{ Consider the cylindrical geometry for now}.
By $r$ and $R$ we denote the transverse coordinate and the nominal cylinder radius,
    respectively.  
By ``modes'' we broadly refer to the solutions to 
   the pertinent eigenvalue problems (EVPs) defined on the open domain $[0, \infty)$,
   { with a mode jointly characterized by 
        some ``mode frequency'' and ``mode function''.
A mode is not necessarily an eigenmode despite the
   broad term ``eigenvalue problem''. 
We reserve the term ``eigenmodes'' for those that qualify as such,
   deeming ``eigenfrequencies'' and ``eigenfunctions''
   as applicable only to eigenmodes
   \footnote{In our context, the distinction between modes and eigenmodes is necessary only for EVPs defined on an open domain given the subtleties with the boundary condition (BC) at infinity. The specification of BCs is straightforward if a closed domain is adopted, 
   all solutions qualifying as eigenmodes in terms of orthogonality and completeness
   \citep[e.g.,][]{1989CourantHilbert}.}.}  
Eigenvalues are necessarily real-valued { in ideal MHD} 
   (see e.g., \citealt{2007PhPl...14e2101A,2014ApJ...789...48O,2015ApJ...806...56O} hereafter
   \citetalias{2007PhPl...14e2101A,2014ApJ...789...48O,2015ApJ...806...56O}, respectively)
   \footnote{
   This does not contradict the notion of ideal quasi-modes
   (see Chapter~10 in \citealt{2019CUP_goedbloed_keppens_poedts};
    also Section~3 in \citealt{2014ApJ...788....9G}).
   Take resonantly damped kink motions in a cylinder where the density
       varies continuously over some transition layer (TL) around the boundary.
   For thin TLs, \citet{2015ApJ...803...43S} showed that the coordinated kink motions
       can maintain their global character, making their temporal variation
       describable by the complex frequencies         
       computed with either the dissipative eigenmode approach 
       \citep[e.g.,][]{1991PhRvL..66.2871P}
       or the ideal quasi-mode approach \citep[e.g.,][]{2013ApJ...777..158S}.
   However, the meaning of quasi-modes becomes less and less clear when
       the TL width increases.}.
Eigenfunctions can then be made real-valued, enabling 
   { an eigenmode to be classified}
   as either evanescent or oscillatory.
By ``evanescent eigenmodes'' we refer to those for which the eigenfunctions
   in the exterior are evanescent somewhere, whereas by ``oscillatory eigenmodes'' we refer to those whose eigenfunctions are oscillatory throughout
   the exterior.        
Overall, two groups of EVPs have been extensively examined in the literature,
   the difference being in the treatment of the nominal outer boundary ($r\to \infty$).
No boundary condition (BC) is specified in one group (``EVP open noBC'' hereafter),
   whereas the BC for { the other group (``EVP open noIC'') is that no
   incoming waves} are allowed. 

Now consider the simplest case where the equilibrium density takes 
   a step profile, evaluating to $\rhoi$ ($\rhoe$) where $r<R$ ($r>R$).
Let $\vai$ and $\vae$ denote the \Alf\ speeds evaluated with $\rhoi$ and $\rhoe$,
   respectively.    
A series of cutoff axial wavenumbers $\{k_{{\rm cutoff}, m}\}$ 
   are relevant, satisfying $k_{{\rm cutoff}, 1}<k_{{\rm cutoff}, 2}<\cdots$
   \citep[e.g.,][]{1984ApJ...279..857R,2014ApJ...781...92V}.   
By relevant we mean that a discrete set of $m$ evanescent eigenmodes is allowed 
   when $k>k_{{\rm cutoff}, m}$, 
{   their real-valued frequencies being consistently lower than the critical frequency
   $\omgcrit=k\vae$. 
The focus of the classic study by \citet{1983SoPh...88..179E},
   this set} is actually the evanescent part of the solutions to
   both ``EVP open noBC'' \citepalias{2015ApJ...806...56O} 
   and { ``EVP open noIC''} \citep[e.g.,][]{1986SoPh..103..277C,2015ApJ...812...22C}.
Note that this evanescent set is empty when $k<k_{{\rm cutoff}, 1}$.
Note also that evanescent eigenmodes are customarily referred to as
   ``trapped modes'' \citep[e.g.,][]{1983SoPh...88..179E,2005LRSP....2....3N,2020ARA&A..58..441N}.
The key difference between { ``EVP open noIC''} and ``EVP open noBC''
   is then that, for arbitrary $k$,
   an infinity of { discrete leaky modes (DLMs) with complex-valued frequencies} arise for the former 
   \citep[e.g.,][]{1978SoPh...58..165M,1982SoPh...75....3S,1986SoPh..103..277C}, 
   whereas a continuum of oscillatory eigenmodes { with real-valued frequencies} arises for the latter
   (\citetalias{2015ApJ...806...56O}; see also \citetalias{2007PhPl...14e2101A,2014ApJ...789...48O}).
Now focus on the situation where $k< k_{{\rm cutoff}, 1}$, which
   tends to be the only situation relevant for AR loops given
   their large length-to-radius ratios and mild density contrasts 
   \citep[][]{2004ApJ...600..458A}.     
A multitude of { initial value problem (IVP) studies}
   have established that DLMs, while not eigenmodes per se, 
   may account for a substantial { fraction of the system evolution}
   \citep[e.g.,][]{2007SoPh..246..231T,2012ApJ...761..134N,2015ApJ...812...22C,2016SoPh..291..877G}
   \footnote{See, e.g., Figure~3 in \citet{2007SoPh..246..231T} for the temporal
    evolution of a sausage perturbation with $k<k_{\rm cutoff, 1}$. 
   Figures~6 and~7 therein further suggest that the leaky phase occurs only
   in a brief interval for the temporal evolution of kink perturbations, which are beyond
   the scope of this study. 
   }. 
On the other hand, it is well known that 
   $P^{\rm DLM} \coloneqq 2\pi/\Re\omega^{\rm DLM} < 2.6R/\vai$ 
   and $\tau^{\rm DLM}/P^{\rm DLM} \coloneqq |\Re\omega^{\rm DLM}/\Im\omega^{\rm DLM}|/(2\pi) \sim (\rhoi/\rhoe)/\pi^2$, 
   with $P^{\rm DLM}$ and $\tau^{\rm DLM}$ 
   { representing the period and damping time,
   respectively \citep[e.g.,][]{2007AstL...33..706K}}.
{ Sausage perturbations are therefore difficult} 
   to observe in AR loops given their short periodicities and rapid attenuation. 
   
Do sausage perturbations remain difficult
   to observe in AR loops for other profiles of the equilibrium density $\rho_0$? 
This question is sensible { given the intricacies
   for quantifying the transverse density profiles}
   with, say, EUV measurements 
   \citep[see e.g.,][]{2003ApJ...598.1375A,2017A&A...600L...7P,2017A&A...605A..65G}. 
We proceed with {  the choice} 
   labeled ``outer $\mu$'' in \citet{2017ApJ...836....1Y}, 
   where $\rho_0$ evaluates to $\rhoi$ when
   $r<R$ but otherwise reads $\rhoe+(\rhoi-\rhoe)(r/R)^{-\mu}$.
Here $R$ { is some} nominal cylinder radius,   
   and { $\mu\ge 1$ represents some steepness parameter}. 
\citet[][hereafter \citetalias{2015ApJ...810...87L}]{2015ApJ...810...87L}
   were the first to examine an EVP of the ``EVP open noBC'' type 
   for this family of density profiles
   \footnote{The theoretical analyses in \citetalias{2015ApJ...810...87L} are not specific to the ``outer $\mu$'' family, for which our discussions here
   follow directly from the more general results therein.},  
   focusing on the behavior of the solutions at infinity 
   to single out evanescent eigenmodes.
{ Two distinct regimes were shown to arise as dictated by $\mu$}.
{ A} series of cutoff axial
   wavenumbers $\{k_{{\rm cutoff}, m}\}$ exist when $\mu \ge 2$, 
   satisfying $k_{{\rm cutoff}, 1} \le k_{{\rm cutoff}, 2} \le \cdots$
   \citep[see also][]{2018ApJ...855...53L}.  
The behavior of evanescent eigenmodes is therefore qualitatively the same
   { as in the step case}.
More interesting is the regime where $\mu<2$, for which \citetalias{2015ApJ...810...87L}
   showed that cutoff wavenumbers become irrelevant { in that}
   evanescent eigenmodes are permitted for arbitrary $k$, regardless of
   { $\mu$ and $\rhoi/\rhoe$}.
Furthermore, { the frequencies of evanescent eigenmodes
   tend to the critical frequency $\omgcrit = k\vae$ when $kR \to 0$.}   
{ Note that this $\omgcrit$ translates into  
   a period $2L/\vae$} for axial fundamentals with $L$ being
   the loop length.    
{ \citetalias{2015ApJ...810...87L} (see also \citealt{2019MNRAS.488..660L})
   accepted the dominance of evanescent eigenmodes, arguing that   
   sausage perturbations may be invoked to account for 
   flare QPPs with quasi-periods of both $\bigO(R/\vai)$ and $\bigO(L/\vae)$.
Now focus again on AR loops.  
One deduces that sausage perturbations may be readily captured in observations 
   for AR loops with $\mu<2$, given the likely dominance of evanescent eigenmodes
   and the long periodicities.}
However, this deduction was not seen in the IVP study 
   by \citet[][\citetalias{2022ApJ...928...33L}]{2022ApJ...928...33L}.
{ For $\mu>2$ and $\mu<2$ alike, 
   the sausage perturbations are subject to rapid attenuation, and their
   periodicities are consistently $\bigO(R/\vai)$ before
   the signals become too weak to be of observational relevance.} 
   
This study aims to examine sausage perturbations in a slab configuration with
   the ``outer $\mu$'' type of equilibrium density profiles. 
Our motivations are threefold.
Firstly, sausage perturbations in slab configurations prove
   to be relevant for oscillatory signals observed in 
   { a broad range} of passbands, some examples being those in EUV \citep[e.g.,][]{2005A&A...430L..65V}
   and radio \citep[e.g.,][]{1983Natur.305..688R,2011A&A...529A..96K,2012A&A...537A..46J,2021MNRAS.505.3505K}. 
Secondly, the mathematical treatments are less involved
    for slab than for cylindrical configurations, making it 
    { easier to concentrate on conceptual understandings}.
Thirdly, the EVP study by 
    \citet[][\citetalias{2015ApJ...801...23L} hereafter]{2015ApJ...801...23L}
    demonstrated that the behavior of cutoff wavenumbers in slab configurations
    is qualitatively identical to its cylindrical counterpart
    \footnote{By this assertion we mean that cutoff axial wavenumbers exist only when $\mu \ge 2$, for slab and cylindrical geometries alike     
    \citep[see][Figure~8]{2018ApJ...855...53L}.
    The mode functions, however, differ quantitatively in their dependencies
        on the transverse coordinate.
    Consider trapped sausage modes in coronal equilibria with step density profiles.
    The transverse velocity in the exterior follows $K_1(\kappae r)$
        ($\Exp{-\kappae x}$) for cylindrical (slab) equilibria, with 
        $\kappae = \sqrt{k^2 - \omega^2/\vae^2}$ and $K_1$
        the modified Bessel function of the second kind \citep[e.g.,][]{2019CUP_Roberts}.
    The former profile tends to drop off more rapidly with distance.
    We note by passing that the same difference in the drop-off rate takes place 
        for trapped kink modes as well
        (see e.g., \citealt{1992SoPh..138..233G,2007SoPh..246..213A,2021ApJ...921...29H,2021SoPh..296...95Y} for the connection between the two geometries).
    }, 
    thereby warranting
    a slab version of the IVP study by \citetalias{2022ApJ...928...33L}.           
We nonetheless stress that { this study} is new in the following aspects.
One, we will { consistently solve the IVP} in terms of eigensolutions to
    an EVP defined on a closed domain.
{ This ``Modal Closed'' approach has been employed only in the cylindrical studies 
     on sausage perturbations by \citet{1996ApJ...472..398B} and \citetalias{2022ApJ...928...33L} to our knowledge.}
Two, we will { revisit the IVP study for step density profiles
     by \citetalias{2007PhPl...14e2101A}, offering
     a Fourier-integral-based formal solution rather than 
     following the Laplace transform approach therein.
Key to this formal solution are the eigensolutions to   
     an EVP of the ``EVP open noBC'' 
     type (see \citetalias{2014ApJ...789...48O,2015ApJ...806...56O,2022ApJ...928...33L}).
Our study complements \citetalias{2007PhPl...14e2101A} by further addressing
     the connection between continuum eigenmodes and discrete leaky modes.} 
We will additionally demonstrate how the oscillatory continuum for open systems
    is recovered by the discrete eigenspectrum in our ``Modal Closed'' approach.    
Three, we will work on the more general situation with finite $\mu$, paying special 
    attention to the influence of $\mu$ on the connection between 
    the discrete eigenspectrum and the oscillatory continuum. 

This manuscript is structured as follows.
We formulate the IVP for a laterally open system in 
    Section~\ref{sec_IVP_formu}, where we also detail the ``Modal Closed'' approach.
Section~\ref{sec_step} then largely proceeds analytically to examine
    the IVP for a step density profile.
We move on to numerically examine the more general situation where
    $\mu$ is finite in Section~\ref{sec_finitemu}.      
Section~\ref{sec_conc} summarizes the present study, 
    ending with some concluding remarks.

\section{Problem Formulation}
\label{sec_IVP_formu}
  
\subsection{Equilibrium and Overall Description}
\label{sec_sub_equili}

We work in the framework of pressureless, ideal, MHD throughout.
Let $\rho$, $\vec{v}$, and $\vec{B}$ denote 
   the mass density, velocity, and magnetic field, respectively. 
We denote the equilibrium quantities with a subscript $0$, 
   and assume that the equilibrium is static  ($\vec{v}_0 = 0$).
Working in a Cartesian coordinate system $(x, y, z)$,
   we take the equilibrium magnetic field to be uniform
   and $z$-directed
   ($\vec{B}_0 = B_0 \vec{e}_z$). 
We assume that the equilibrium density ($\rho_0$)
   depends only on $x$
   and is symmetric about $x=0$.
{ Specifically, $\rho_0(x)$ is chosen to be
    the ``outer $\mu$'' family in \citet{2017ApJ...836....1Y},}
\begin{equation}
\label{eq_rho_prof_outermu}
\begin{split} 
& {\rho_0}(x)
   = \rhoe+(\rhoi-\rhoe) f(x), 
\\
&  f(x) 
   =
    \left\{
       \begin{array}{ll}
          1,     						& ~~0 \le x \le R, 			\\[0.1cm]
          \left(x/R\right)^{-\mu}, 	    & ~~x \ge R.
       \end{array} 
    \right.
\end{split}   
\end{equation} 
Here $R$ represents some nominal slab half-width.    
With the subscript ${\rm i}$ (${\rm e}$)
   we denote the equilibrium quantities at $x=0$ ($x\to \infty$). 
The \Alf\ speed is defined by $\va^2 = B_0^2/(\mu_0 \rho_0)$ with $\mu_0$ being
   the magnetic permeability of free space. 
By $\vai$ ($\vae$) we then refer to the \Alf\ speed   
   at the slab axis (infinitely far).
Evidently, the steepness parameter $\mu$ is such that
   $\rho_0$ ($\va$) decreases (increases) more rapidly with $x$ 
   towards $\rhoe$ ($\vae$) when $\mu$ increases. 

Let the subscript $1$ denote small-amplitude perturbations.
We neglect both out-of-plane propagation ($\partial/\partial y =0$)   
   and the out-of-plane components of vectorial quantities.
Consequently, only $v_{1x}$, $B_{1x}$, and $B_{1z}$ survive
   in linearized, pressureless, ideal MHD. 
With axial standing waves in mind, we take coronal slabs to be 
   bounded by the planes $z=0$ and $z=L$, and adopt
   the following ansatz
\begin{equation}
\label{eq_Fourier_ansatz}
\begin{split}
  v_{1x} (x, z,  t) 
& = \hat{v}  (x, t) \sin(kz), 
\\
  B_{1x} (x, z, t) 
&= \hat{B}_x(x, t) \cos(kz), 
\\
  B_{1z} (x, z, t) 
&= \hat{B}_z(x, t) \sin(kz), 
\end{split}
\end{equation}     
   where $k = n \pi/L$ is the quantized 
   axial wavenumber ($n=1, 2, \cdots$).
One finds that $\hat{B}_x$ and $\hat{B}_z$ are connected to $\hat{v}$ via
\begin{eqnarray}
   \dfrac{\partial \hat{B}_{x}}{\partial t}
&=&  
   k B_0 \hat{v}, 
	      	\label{eq_linMHD_Farad_comp_x_Fourier} 
\\
         \dfrac{\partial \hat{B}_{z}}{\partial t}
&=&  
   -B_0  \dfrac{\partial \hat{v}}{\partial x}. 
	      \label{eq_linMHD_Farad_comp_z_Fourier} 
\end{eqnarray}
We assume that 
\begin{eqnarray}
   \hat{B}_x (x, t=0) = \hat{B}_z (x, t=0) = 0 \label{eq_IC_B} 
\end{eqnarray}
   without loss of generality.
The initial velocity perturbation is taken to be 
\begin{eqnarray}
u(x) \coloneqq  \hat{v}(x, t=0) =
    \left\{
       \begin{array}{ll}
          \vai \sin^3(\pi x/\Lambda),   & \quad 0 \le x \le \Lambda, 			\\[0.1cm]
          0, 	    			   & \quad x \ge \Lambda,
       \end{array} 
    \right.
 \label{eq_u}
\end{eqnarray}    
    which is localized within $x=\Lambda$
    and prescribed to be sufficiently smooth with a magnitude arbitrarily
    set to be $\vai$.
Only sausage perturbations arise given that $u(x=0)=0$
    and that $\rho_0(x)$ is even. 
{ It suffices to consider only the half-plane $x\ge 0$.}     

\subsection{Initial Value Problem and Energetics}
\label{sec_sub_probformu}
We retain only $\hat{v}$ to formulate our IVP for convenience.
\begin{IVP} 
\label{ivp_mu_open}
Solutions are sought for the following equation 
   \begin{eqnarray}
   \displaystyle 
      \frac{\partial^2 \hat{v}}{\partial t^2}
   &=&
      \va^2(x)
      \left(
          \frac{\partial^2 \hat{v}}{\partial x^2}
        - k^2 \hat{v}    
      \right),  \label{eq_v2nd_final}    
   \end{eqnarray}
   { subject to the} initial conditions (ICs)
\begin{eqnarray}
\hat{v}(x, t=0) = u(x), \quad  
\frac{\partial \hat{v}}{\partial t} (x, t=0) = 0,
\label{eq_v2nd_IC}
\end{eqnarray}
   together with the boundary condition (BC)
\begin{eqnarray}
\hat{v}(x=0, t) = 0, \label{eq_FD_BC}
\end{eqnarray}   
   on a domain spanning from $x=0$ to $\infty$.
\end{IVP}
Note that no BC is required at $x\to\infty$.
Note further that the IC for $\partial\hat{v}/\partial t$
   is based on the IC~\eqref{eq_IC_B}, from which
   $\hat{B}_x(x,t)$ and $\hat{B}_z(x,t)$ can be found by
   integrating Equations~\eqref{eq_linMHD_Farad_comp_x_Fourier}
   and \eqref{eq_linMHD_Farad_comp_z_Fourier} 
   over time
   once $\hat{v}(x,t)$ is known.

It proves necessary to examine the associated energetics.
Let $\mathcal{V}$ refer to a volume that is of unit length in the $y$-direction
   and spans the rectangle $[0, x]\times[0, L]$ in the $x-z$ plane. 
An energy conservation law is then readily derived, reading
\begin{eqnarray}
     E_{\rm tot}(x, t) - E_{\rm tot}(x, t=0) 
  = -F(x, t),
  	\label{eq_ConsInt_Etot}
\end{eqnarray}           
   where 
\begin{eqnarray}
     E_{\rm tot}(x, t)    
&=&  
     \dfrac{L}{2} \int_{0}^{x} dx'
       \left\{
           \dfrac{1}{2}\rho_0 (x') \hat{v}^2(x', t)
         + \dfrac{1}{2\mu_0}
                   \left[ \hat{B}_x^2(x', t)
                        + \hat{B}_z^2(x', t)
                   \right] 
       \right\}~, 
     \label{eq_def_ener_Etot} 
\\
	F(x, t) 
&=& 
    \dfrac{L}{2} \int_{0}^{t} dt'
	\left[\hat{p}_{\rm T}(x, t') \hat{v} (x, t')\right]~.   
	\label{eq_def_ener_F}           
\end{eqnarray} 
Here $\hat{p}_{\rm T} = B_0 \hat{B}_z/\mu_0$ represents
    the Eulerian perturbation of total pressure, and  
    the terms in the square parentheses in 
        Equation~\eqref{eq_def_ener_F} stem from the lateral component
        of the Poynting vector. 
Furthermore, the common factor $L/2$ is retained to ensure that
    $E_{\rm tot}(x, t)$ represents the instantaneous total energy in $\mathcal{V}$,
    while $F(x, t)$ represents the cumulative energy loss from $\mathcal{V}$.

\subsection{Solution Methods}
\label{sec_sub_method}
We choose to solve IVP~\ref{ivp_mu_open} with two independent methods,
    one being a finite-difference (FD) approach,
    and the other being a modal approach involving eigensolutions
    to { an EVP on a finite domain.}
The modal solutions turn out to be more physically insightful.
The FD approach, however, is orders-of-magnitude less time-consuming.
Two situations then arise where we present the FD solutions.
First, different approaches may need to be cross-validated.
Second, parametric studies are needed or
    the FD approach is more convenient to work with.

\subsubsection{The Finite-Difference Approach}
\label{sec_sub2_FDapproach}
Our FD scheme is second-order accurate in both space and time, 
    and is essentially identical to its cylindrical counterpart 
    in \citetalias{2022ApJ...928...33L}.
A uniform grid with spacing $\Delta x = 0.01~R$ is employed for simplicity. 
The timestep is specified as $\Delta t = c \Delta x/\vae$, 
    where the Courant number $c$ is chosen to be $\sim 0.4$ to maintain
    numerical stability. 
We ensure that no difference can be discerned when 
    $\Delta x$ or $c$ varies. 
More importantly, we make sure that our FD solutions remain the same
    when the outer boundary is placed at larger distances.

\subsubsection{The ``Modal Closed" Approach}
\label{sec_sub2_ModalApproach}
Our modal approach, labeled ``Modal Closed'' hereafter, 
    relies on the following EVP. 
\begin{EVP} 
\label{evp_mu_closed}
Nontrivial solutions are sought for the following equation 
\begin{equation}
   \omega^2 \breve{v} 
= -\va^2(x)
   \left( 
	   \dfrac{d^2}{dx^2}\breve{v} 
 			      - k^2 \breve{v}
   \right)
\equiv \mathcal{L}\breve{v}, 
\label{eq_modal_EVP_govE}
\end{equation}
   defined on a domain of $[0, d]$ and { subject to} the BCs
\begin{equation}
\breve{v} (x=0) = \breve{v} (x=d) = 0.
\label{eq_modal_EVP_BC}
\end{equation} 
\end{EVP}
Equation~\eqref{eq_modal_EVP_govE} is found by replacing $\hat{v}$ with
    $\Re[\breve{v}(x) \exp(-i\omega t)]$ in
    Equation~\eqref{eq_v2nd_final}.
One readily recognizes that the operator $\mathcal{L}$ is Hermitian
    under the definition of the scalar product 
\begin{eqnarray}
\label{eq_def_innerprod_closed}
    \innerpClosed{U}{V}
\equiv    
    \int_{0}^{d}
    U^*(x)V(x) \rho_0(x) dx,
\end{eqnarray}    
    where the asterisk represents complex conjugate.  
{ The superscript $(d)$ is meant to
	emphasize that EVP~\ref{evp_mu_closed}
	   is defined on a closed domain}.
The following general properties (GPs) then follow from general theory.
\begin{GP}
\label{GP_omg2seq}
The countable infinity of positive eigenvalues 
    $\{\omega^2_{l}\}$ form a monotonically increasing sequence
    with respect
        to the mode number $l = 1, 2, \cdots$.
\end{GP}
\begin{GP}
\label{GP_omg2onDomainSize}
The $l$-th eigenvalue for a domain never exceeds 
    the $l$-th eigenvalue for its subdomain.
\end{GP}
\begin{GP}
\label{GP_func_real}
The $l$-th eigenfunction $\breve{v}_l(x)$ can be made
    real-valued, possessing $l-1$ nodes inside the domain.
\end{GP}
\begin{GP}
\label{GP_func_ortho}
The set $\{\breve{v}_l(x)\}$ is complete, satisfying
    the orthogonality condition  
\begin{eqnarray}
   \innerpClosed{\breve{v}_l}{\breve{v}_{l'}} = \innerpClosed{\breve{v}_l}{\breve{v}_{l}}\delta_{l, l'},
\end{eqnarray}    
   where $\delta_{l, l'}$ denotes the Kronecker delta.
\end{GP}

We specialize to our equilibrium density
   profile~\eqref{eq_rho_prof_outermu} from here onward. 
To start,     
   the eigenfrequency $\omgl$ can be expressed formally as
\begin{equation}
\label{eq_EVP_omgl_formal}
    \dfrac{\omgl R}{\vai} 
=  {\mathcal{F}}_l(\rhoi/\rhoe, \mu; k R; d/R).
\end{equation} 
We see $\omgl$ as positive without loss of generality. 
It now proves helpful to put Equation~\eqref{eq_modal_EVP_govE}
   in a  \Schrod\ form as
\begin{equation}
\label{eq_EVP_schro}
\dfrac{d^2 \breve{v}}{d x^2} + Q(x) \breve{v}  = 0, 	
\end{equation}
where the coefficient $Q(x)$ is defined through a potential $V(x)$ as
\begin{equation}
\label{eq_def_QV}
V(x)  = k^2 \va^2(x), \quad 
Q(x)  = \dfrac{\omega^2 - V(x)}{\va^2(x)}
      = \dfrac{\omega^2}{\va^2(x)} - k^2.
\end{equation}
The following specific properties (SPs) of EVP~\ref{evp_mu_closed} then ensue,
    where by ``specific'' we mean that the properties are more or less 
    specific to our equilibrium density profile. 
Nonetheless, they hold regardless of the specific values of
    $[\rhoi/\rhoe, \mu, kR, d/R]$.    
\begin{SP}
\label{SP_omglowerbound}
The eigenfrequency $\omgl$ for any $l$ is bound to exceed $k\vai$. 
\end{SP}    
\begin{SP}
\label{SP_omglhigh}
High-frequency eigenmodes with $\omgl^2 \gg \omgcrit^2=(k\vae)^2$ are always permitted, 
   following 
\begin{equation}
\label{eq_omglhigh}
\omgl \approx l\pi (\vae/d), \quad 
\text{provided that}~ 
l^2 \gg  (k d/\pi)^2.
\end{equation}
\end{SP}    
\begin{SP}
\label{SP_OsciEva}
The eigenfunction $\breve{v}_l$ is oscillatory where $\omega_l^2>V(x)$ but evanescent
    otherwise. 
\end{SP}    
We note that SP~\ref{SP_omglowerbound} can be justified by the heuristic argument that 
   if $\omgl< k\vai$, then $\breve{v}_l$ necessarily cusps somewhere in the domain and
   therefore the continuity requirement for $d\breve{v}_l/dx$ is violated.
The same argument was first offered
   by \citet{1996ApJ...472..398B} in cylindrical geometry.
Likewise, a similar version of SP~\ref{SP_omglhigh} holds in
   cylindrical geometry as detailed in
   Appendix~B in \citetalias{2022ApJ...928...33L}.
Following the WKB treatment therein yields Equation~\eqref{eq_omglhigh} for
   the Cartesian geometry, and 
   the range of validity turns out to be less restrictive.
Important for { this study} is SP~\ref{SP_OsciEva}, 
   with which we classify the eigenmodes
   for a finite $d/R$ into two groups by the spatial behavior of the corresponding
   eigenfunctions.
An eigenmode, labeled by $l$, is said to be evanescent
   if its eigenfunction is evanescent
   somewhere in the domain.
In contrast, an eigenmode is deemed oscillatory 
   if its eigenfunction is oscillatory throughout.
{ Now that $\va(x)$ increases monotonically}
   toward $\vae$ when $x$ increases,    
   an eigenmode is necessarily oscillatory 
   for a given $d/R$ if $\omgl > \omgcrit = k\vae$ but evanescent 
   when the opposite is true.       
Furthermore, 
   if an eigenmode transitions from an oscillatory 
      to an evanescent one when $d/R$ exceeds some critical value,
      then property GP~\ref{GP_omg2onDomainSize} dictates that
      it will remain an evanescent eigenmode for larger still $d/R$.         
   
{    
We now connect EVP~\ref{evp_mu_closed} to the general EVP analysis
   in \citetalias{2015ApJ...801...23L}. 
Put briefly, the EVP therein belongs to the ``EVP open noBC'' category
   because the authors analyzed the
   behavior of mode functions at infinity rather than specifying some BC beforehand. 
The following deductions can be made by 
   specializing to our profile $f(x)$ (see Equation~\eqref{eq_rho_prof_outermu}).
Consider the situation $\mu \ge 2$ first. 
Equations~(12) to (16) in \citetalias{2015ApJ...801...23L} mean that
   there exist a series of cutoff axial wavenumbers
   $k_{{\rm cutoff}, m}$ with $m=1, 2, \cdots$.
\citet{2018ApJ...855...53L} further showed that $k_{{\rm cutoff}, m}$ is expressible as
\begin{equation}
\label{eq_def_kcut_muprofile}
  k_{{\rm cutoff}, m}R = \dfrac{g_m}{\sqrt{\rhoi/\rhoe-1}} 
\quad 
  (m=1, 2, \cdots),
\end{equation}
   where the $\rhoi/\rhoe$-independent coefficient $g_m \ge 1/2$.}
One then deduces that no evanescent eigenmodes are permitted if $k<k_{{\rm cutoff},1}$,
   with the inequality bound to hold when
\begin{equation}
\label{eq_def_kInEq4osci}
kR  < \dfrac{1/2}{\sqrt{\rhoi/\rhoe-1}}.
\end{equation}   
Now consider the situation where $\mu<2$, for which Equations~(18) and (19)
   in \citetalias{2015ApJ...801...23L} lead to the peculiar behavior for 
   { cutoff wavenumbers to disappear altogether}.
However, what needs to be treated with caution is not this peculiarity but 
   the assertion that ``all fast modes are trapped'' for any $k$ and hence
   ``such plasma slab configurations behave as perfect waveguides''
   (page 3, \citetalias{2015ApJ...801...23L}) in view of our EVP~\ref{evp_mu_closed}. 
Our point is, an infinity of oscillatory eigenmodes are ensured by 
    property SP~\ref{SP_omglhigh} regardless of $d/R$,
    making them difficult to disappear altogether
    in the limit $d/R \to \infty$.
More importantly, the solution to IVP~\ref{ivp_mu_open} at any location
    can be expressed in terms of 
    the eigensolutions to EVP~\ref{evp_mu_closed} provided that
    a timeframe of validity is specified. 
Ideally, this confusion should be resolved with
    a theoretical understanding of the spectrum associated with the EVP examined in
    \citetalias{2015ApJ...801...23L}. 
However, we argue that there is no need to do so as long as the solutions
    to IVP~\ref{ivp_mu_open} can be understood with the aid of 
    the textbook properties of EVP~\ref{evp_mu_closed}.
In particular, the ``Modal Closed'' approach allows to make clear 
    what is meant by the irrelevance of cutoff wavenumbers. 
{ We therefore restrict ourselves to the situation where 
    Inequality~\eqref{eq_def_kInEq4osci} holds.}

{ 
Some remarks remain necessary on the role of $\mu$ for EVP~\ref{evp_mu_closed}
    when $k<k_{{\rm cutoff}, 1}$.
To start, property GP~\ref{GP_omg2onDomainSize}       
    follows from Theorem~3 in \citet[][Chapter VI]{1989CourantHilbert}, 
    where it was noted that the $l$-th eigenvalue for a domain is in fact smaller
    than the $l$-th eigenvalue for any of its subdomains.     
When $\mu \ge 2$, one then directly deduces that 
    EVP~\ref{evp_mu_closed} allows only oscillatory eigenmodes
    for any finite domain, because evanescent eigenmodes are prohibited
    for a truly open domain.  
For $\mu < 2$, on the other hand, some concern may arise when
    we compute evanescent eigenmodes for EVP~\ref{evp_mu_closed}.
Let $l$ refer to an evanescent eigenmode, and let $\omgl(d)$
    be its eigenfrequency evaluated with a domain of size $d$. 
The stronger version of property GP~\ref{GP_omg2onDomainSize}
    then dictates that the eigenfrequency $\omgl$ depends on the domain size,
    meaning specifically that $\omgl(d') < \omgl(d)$ for any $d'>d$.
In practice, however, we need to quote some domain-independent eigenfrequency
    for definiteness (e.g., Figure~\ref{fig_D1related}).
This discrepancy is nonetheless only apparent, because what we mean
    by domain-independent is that $\omgl(d')$ and $\omgl(d)$ 
    are equal within the accuracy of our numerical solver when $d$
    is sufficiently large. 
}


With the above preparations, { the ``Modal Closed'' solution} 
   to IVP~\ref{ivp_mu_open} writes
\begin{equation}
\label{eq_modal_formalSol}
\begin{split}
&  \vhatd (x, t)
  = \sum\limits_{l = 1}^{\infty} 
        c_l  \breve{v}_l(x) \cos(\omega_l t), 
\\
& 0 \le x \le d, \quad         
  t \le \int_{\Lambda}^{d}  \dfrac{dx}{\va(x)}.
\end{split}
\end{equation}
Here the coefficient $c_l$ measures the contribution
    from the $l$-th eigenmode,
\begin{eqnarray}
  c_l 
= \dfrac{\innerpClosed{u}{\breve{v}_l}}
        {\innerpClosed{\breve{v}_l}{\breve{v}_l}}.
  \label{eq_modal_coef}
\end{eqnarray}
Despite the superscript $(d)$, we stress that $\vhatd(x, t)$ itself 
   does not depend on $d$ as long as the 
   outermost edge of the perturbation has not reached $x=d$. 
The relevant timeframe of validity is explicitly 
   specified in Equation~\eqref{eq_modal_formalSol}.
   
Two steps are involved for { numerically evaluating
    the  ``Modal Closed'' solution}.
{ 
Firstly, we follow \citetalias{2022ApJ...928...33L} to formulate and solve 
    EVP~\ref{evp_mu_closed} with the general-purpose finite-element code 
    PDE2D \citep{1988Sewell_PDE2D},
    which was introduced into the solar context by \citet{2005ApJ...618L.149T}
    \footnote{PDE2D has subsequently been extensively employed in solar wave studies
        for addressing both IVPs  \citep[e.g.,][]{2006ApJ...642..533T,2011A&A...531A.167S,2016ApJ...818..128O}
        and EVPs
        \citep[e.g.,][]{2007A&A...466.1145A,2011A&A...533A..60A,2021ApJ...908..230C}.
    Recently, it found application to the construction of  
        magnetohydrostatic equilibria as well \citep{2022A&A...660A.136T}.}}.     
As output we have the sets $\{\omgl\}$ and $\{\breve{v}_l\}$ 
    for a specified combination $[\rhoi/\rhoe, \mu; k R; d/R]$.
Secondly, we evaluate the coefficient $c_l$ and
    then the modal solution $\vhatd$
    with Equations~\eqref{eq_modal_coef} and \eqref{eq_modal_formalSol},
    respectively.
{ We ensure that a sufficient number of eigenmodes 
    are incorporated in the sum}.     

\subsection{Parameter Specification}
\begin{figure}
\centering
\includegraphics[width=.85\columnwidth]{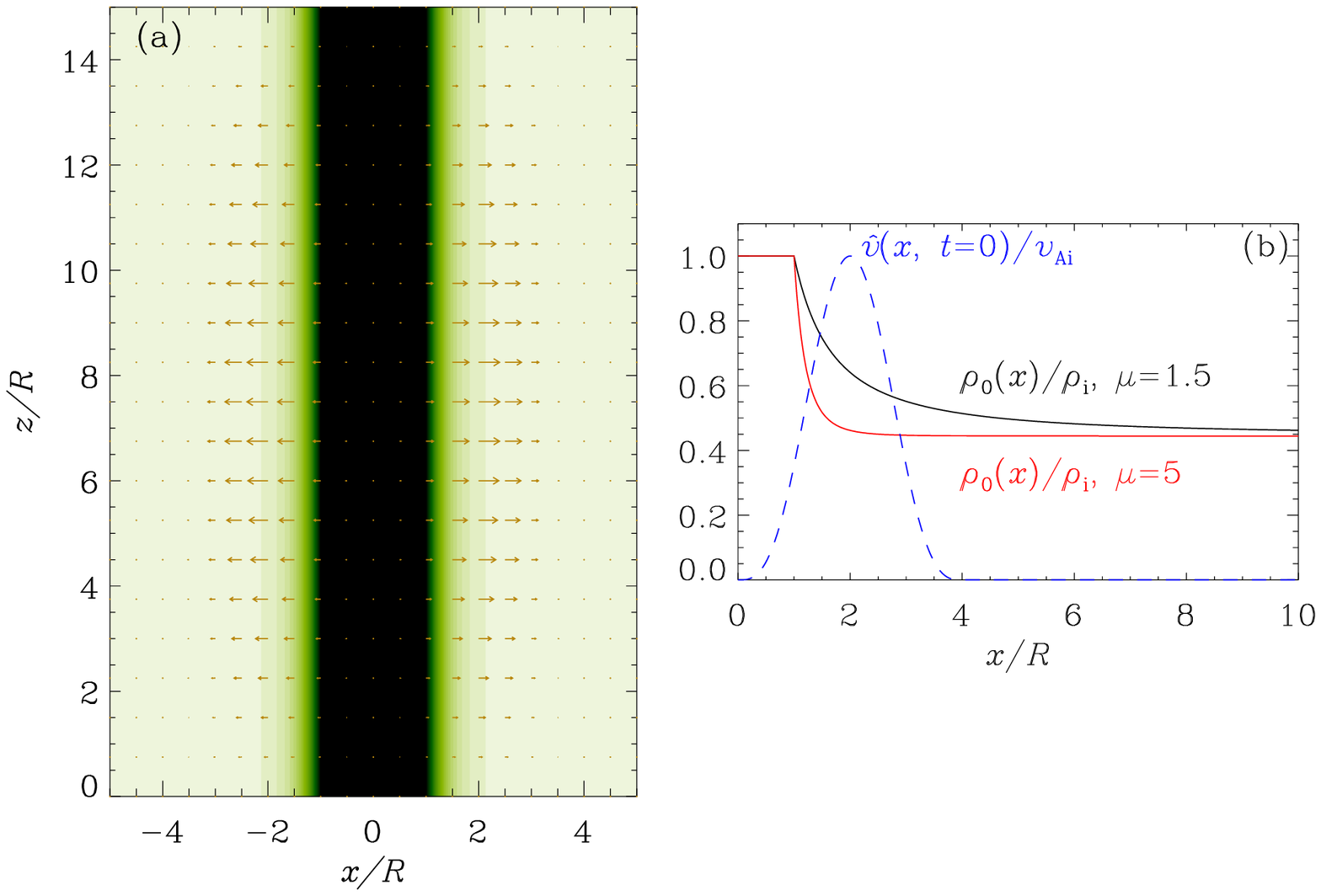}
\caption{
(a) Two-dimensional (2D) representation of the 1D initial value problem
    (IVP~\ref{evp_mu_closed}). 
Plotted are the $x-z$ distributions of the equilibrium density
    (the filled contours)
    and the initial velocity field (the arrows).
The examined slabs possess a nominal half-width $R$.    
Axial fundamentals are ensured by the $z$-dependence
    of the initial perturbation.
(b) Illustrative transverse profiles of the initial perturbation
    ($\hat{v}$, the blue dashed curve) 
    and the equilibrium density ($\rho_0$, the solid curves),
    both involved in IVP~\ref{ivp_mu_open}.
The density profile follows the outer $\mu$ prescription in 
    Equation~\eqref{eq_rho_prof_outermu}, with
    the density contrast $\rhoi/\rhoe$ chosen to be $2.25$. 
Two steepness parameters are plotted, namely 
        $\mu = 1.5$ (the black curve)
    and $\mu = 5$ (red).
The initial perturbation is described by Equation~\eqref{eq_v2nd_IC},
    with the spatial extent $\Lambda$ chosen to be $4R$ here.     
 }
\label{fig_EQprofile} 
\end{figure}

Let us start by recalling that the solution to IVP~\ref{ivp_mu_open}
    is determined by the combination $[\rhoi/\rhoe, \mu; kR; \Lambda/R]$.
We see the steepness parameter $\mu$ as the primary adjustable, and occasionally
    adjust the spatial extent $\Lambda$ { as well}.
The combination $[\rhoi/\rhoe, kR]$ is fixed at $[2.25, \pi/15]$ throughout. 
We examine only axial { fundamentals, for which
    the adopted $kR$}  
    translates into a value of $L/R = 15$ for the ratio of the axial slab length
    to the nominal half-width. 
Note that this pair of $[\rhoi/\rhoe, L/R]$ is reasonable for AR loops, although they
    somehow lie close to the lower ends of the observed ranges
    deduced from EUV measurements
    \citep[e.g.,][]{2004ApJ...600..458A,2007ApJ...662L.119S}.
Note further that this pair 
    yields a $(kR) \sqrt{\rhoi/\rhoe-1}$ of $0.23$, and hence
    ensures Inequality~\eqref{eq_def_kInEq4osci}.
Figure~\ref{fig_EQprofile}a offers a two-dimensional (2D) representation
    of IVP~\ref{ivp_mu_open}
     by presenting the 
     $x-z$ distributions of the equilibrium density (the filled contours)
     and the initial velocity field (the arrows).
The equilibrium density is further plotted against the transverse coordinate
     in Figure~\ref{fig_EQprofile}b, where
     two values are examined for the steepness parameter $\mu$ as labeled.
In addition, the $x$-profile for the initial perturbation 
     is also shown, with $\Lambda$
     chosen to be $4R$ for illustrative purposes. 
    
%
%
%
%

\section{Step Density Profile ($\mu=\infty$)}
\label{sec_step}
This section examines IVP~\ref{ivp_mu_open} for the simplest situation where
    the equilibrium density takes a step profile.
Our purposes are threefold.
Firstly, { the only published analytical study on IVP~\ref{ivp_mu_open}} 
    for the present case was due to \citetalias{2007PhPl...14e2101A},
    who adopted a Laplace transform approach.
{ Our Section~\ref{sec_sub_stepFormalSolOpen} will present a new analytical solution
    by directly working with eigenmodes on an open domain
    within the simpler Fourier framework
    \footnote{Only real frequencies are involved in the Fourier approach.
    The Laplace transform approach, on the other hand, requires contour integrals in the complex frequency plane and hence the identification of the branch cuts of the Green's function. 
    However, we follow \citetalias{2007PhPl...14e2101A} to stress that different expressions from different approaches are necessarily equivalent because they describe the same temporal evolution.}.} 
Secondly, in Section~\ref{sec_sub_stepDLM} we will employ our Fourier solution
    to address the connection between the oscillatory continuum modes
    and the much-studied discrete leaky modes, which in some sense 
    makes more apparent the notion raised by \citetalias{2007PhPl...14e2101A}
    that the initial perturbation is of critical importance.
Thirdly, in Section~\ref{sec_sub_step_converge2cont} we will
    address the connection
    between the oscillatory continuum and the discrete spectrum
    associated with EVP~\ref{evp_mu_closed}.     

\subsection{Formal Solution to the IVP in Terms of Eigenmodes on an Open Domain}
\label{sec_sub_stepFormalSolOpen}
This subsection works out the formal solution to IVP~\ref{ivp_mu_open} in terms
    of eigenmodes on an open domain.
The cylindrical version of this approach was presented in 
    Appendix~A.1 of \citetalias{2022ApJ...928...33L}, which in turn
    was based on the Fourier approach detailed by
    \citetalias{2014ApJ...789...48O}
    and 
    \citetalias{2015ApJ...806...56O}.
{ We see the axial wavenumber $k$ as arbitrary in this subsection. 
While derived for a given $k$, the resulting expressions can find
    immediate applications if a continuous distribution of $k$ is involved
    such as happens for wave trains impulsively excited by localized perturbations
    (see \citetalias{2014ApJ...789...48O,2015ApJ...806...56O}; also 
     \citealt{2023MNRAS.518L..57L}).
}

{ 
The EVP of interest in this subsection is of the ``EVP open noBC'' type,
    meaning specifically that it is 
    defined on $[0, \infty)$
    and no BC is specified at infinity.
A series of cutoff wavenumbers exist and are given by}
\begin{eqnarray}
  k_{{\rm cutoff}, m} R 
=  \dfrac{(m-1/2) \pi}{\sqrt{\rhoi/\rhoe-1}}, 
   		\label{eq_step_kcut} 
\end{eqnarray}          
    where $m=1, 2, \cdots$.  
Evidently, $k_{{\rm cutoff}, 1} < k_{{\rm cutoff}, 2} <\cdots$.
{ Defining}
\begin{equation}
\label{eq_step_defkikekappae}
\begin{split}
& \ki^2  
   =   \dfrac{\omega^2 - k^2 \vai^2}{\vai^2}, \\
& \ke^2  
   =   \dfrac{\omega^2 - k^2 \vae^2}{\vae^2}, \\
& \kappa_{\rm e}^2
   = - \dfrac{\omega^2 - k^2 \vae^2}{\vae^2}
   = -\ke^2, 
\end{split}
\end{equation}  
   we note that $\ki^2 > 0$ because $\omega$ always exceeds $k\vai$.
The eigensolutions are then discriminated by the sign
    of $\ke^2$. 
Proper eigenmodes ($\ke^2 <0$) arise when $k$ exceeds $k_{\rm cutoff, 1}$, 
    characterized by a finite set of discrete frequencies. 
Let $j$ label a proper eigenmode, whose eigenfunction then writes
\begin{eqnarray}
\breve{v}_j (x)=  
  \left\{
    \begin{array}{ll}
      - \dfrac{\vai}{\ki R}      \Exp{-\kappae R} \sin(\ki x),   
		        & \quad 0 \le x \le R, 			\\[0.1cm]
        \dfrac{\vai}{\kappae R}  \cos(\ki R)      \Exp{-\kappae x},   
				& \quad x > R.
    \end{array} 
 \right.
 \label{eq_step_open_vproper}
\end{eqnarray}
Evidently, proper eigenmodes are necessarily evanescent. 
{ Equation~\eqref{eq_step_open_vproper} is written to ensure the continuity of
    $d\breve{v}_j(x)/dx$. 
The continuity of $\breve{v}_j(x)$, on the other hand, yields a dispersion
    relation (DR) that dictates the eigenfrequency $\omega_{j}$,}
\begin{eqnarray}
   \kappae \tan(\ki R) + \ki = 0.
   \label{eq_step_open_properDR}
\end{eqnarray}    
{ Improper eigenmodes ($\ke^2 >0$) are relevant regardless of $k$.
They are characterized by a continuous frequency coverage 
   in the range $(k\vae, \infty)$, and hence are oscillatory.
Note that the concept of dispersion relation does not apply.
We then derive the improper eigenfunction $\breve{v}_\omega (x)$ 
    by solving Equation~\eqref{eq_EVP_schro} under the requirements that 
    both $\breve{v}_\omega$ and $d\breve{v}_\omega(x)/dx$ be continuous
    across $x=R$.
The result writes    
}
\begin{eqnarray}
\breve{v}_\omega (x)=  
  \left\{
    \begin{array}{ll}
      -\vai  \dfrac{\ke}{\ki}  \sin(\ki x),   
		        & \quad 0 \le x \le R, 			\\[0.1cm]
      -\vai 
         \left[\Ac \cos(\ke x) + \As \sin(\ke x) \right],   
				& \quad x > R,
    \end{array} 
 \right.
 \label{eq_step_open_vImproper}
\end{eqnarray}        
   where 
\begin{equation}
\label{eq_step_open_ACAS}           
\begin{split} 
& \Ac =  \dfrac{\ke}{\ki} \sin(\ki R) \cos(\ke R) 
                        - \cos(\ki R) \sin(\ke R), \\
& \As =  \dfrac{\ke}{\ki} \sin(\ki R) \sin(\ke R) 
                        + \cos(\ki R) \cos(\ke R).
\end{split}
\end{equation} 
{ We scale $\breve{v}_\omega$ in such a way that it remains regular
    when $\ke$ approaches zero.}
    
{ 
We solve IVP~\ref{ivp_mu_open} by directly superposing
    the proper and improper eigensolutions.
This ``Modal Open'' approach is possible because
    the eigensolutions in the discrete proper set
    and the improper continuum are complete, 
    satisfying the orthogonality condition
\begin{equation}
\label{eq_def_StepOpen_Ortho}
\begin{split} 
&     \innerpOpen{\breve{v}_j}{\breve{v}_{j'}} 
    = \innerpOpen{\breve{v}_j}{\breve{v}_j} \delta_{j,j'},  \\
&     \innerpOpen{\breve{v}_j}{\breve{v}_\omega} 
    = 0,  \\
&     \innerpOpen{\breve{v}_\omega}{\breve{v}_{\omega'}} 
    = q(\omega) \delta(\omega-\omega').
\end{split}
\end{equation} 
Here the scalar product is defined as
\begin{equation}
\label{eq_def_innerprodOpen}
    \innerpOpen{U}{V}
\equiv    
    \int_{0}^{\infty}
    U^*(x)V(x) \rho_0(x) dx,
\end{equation}
    and the superscript $\infty$ emphasizes that the eigensolutions pertain
    to an open domain.
For the proper eigenmodes, it is straightforward to obtain that
\begin{equation}
\label{eq_stepopen_innerPproper}
  \innerpOpen{\breve{v}_j}{\breve{v}_j}
= (\rhoi \vai^2 R) \dfrac{{\rm e}^{-2 \kappae R}}{2} 
  \left[
    \dfrac{1}{\ki^2 R^2}
  + \dfrac{\cos^2(\ki R)}{\kappae R}
        \left( \dfrac{1}{\ki^2 R^2}
             + \dfrac{\rhoe/\rhoi}{\kappae^2 R^2}
        \right)
  \right].    
\end{equation}
For the improper eigenmodes, however, some algebra is necessary
   for evaluating $q(\omega)$ in front of the generalized function 
   $\delta(\omega-\omega')$ in Equation~\eqref{eq_def_StepOpen_Ortho}.
The net result is that    
\begin{equation}
\label{eq_stepopen_qomg}
  q(\omega) 
= (\rhoe \vai^2) 
  \dfrac{\ke \vae^2}{\omega} 
  \dfrac{\pi(A_c^2+A_s^2)}{2}.
\end{equation}    
The ``Modal Open'' solution to IVP~\ref{ivp_mu_open} eventually writes
\begin{equation}
	\label{eq_step_open_SolAna}      
\begin{split}
&    \hat{v}(x, t)
   = \sum\limits_{j = 1}^{J} c_j     
          \breve{v}_j(x)      \cos(\omega_j t) 
   + \int_{k \vae}^{\infty}  S_\omega 
	      \breve{v}_\omega(x) \cos(\omega   t) d\omega,
\\
&   0< x < \infty, \quad 0< t < \infty,	      
\end{split}	
\end{equation}       
   where 
\begin{equation}
\label{eq_step_open_Sol_DefcjSomg}
\begin{split}
& c_j = \dfrac{\innerpOpen{u          }{\breve{v}_j}}
			{\innerpOpen{\breve{v}_j}{\breve{v}_j}}, \\ 
& S_\omega =                	     
	  \dfrac{\innerpOpen{u          }{\breve{v}_\omega}}
	    	{q(\omega)}
	       =
	  \dfrac{\omega\innerpOpen{u          }{\breve{v}_\omega}}
	    	{(\rhoe \vai^2) (\ke \vae^2)(\pi/2)(A_c^2+A_s^2)}.
\end{split}
\end{equation}
We view $S_\omega \breve{v}_\omega(x)$ as
    some local spectral density.
Furthermore, $J$ counts all proper eigenmodes $j$ 
    that satisfy $k> k_{{\rm cutoff}, j}$.
Evidently, $J=0$ if $k< k_{{\rm cutoff}, 1}$.    
}

{ 
Some remarks are necessary.
To start, proper eigenmodes are the familiar ``trapped modes'', 
    with Equations~\eqref{eq_step_kcut} to \eqref{eq_step_open_properDR}
    well documented in, say,  Chapter~5 of \citet{2019CUP_Roberts}.
Somehow less familiar is the improper continuum $(k\vae, \infty)$, 
    which arises due to the absence of a BC at infinity. 
However, the ``Modal Open'' approach itself is actually well established,
    with Equation~\eqref{eq_stepopen_qomg} resulting essentially
    from the techniques developed in the cylindrical study of \citetalias{2014ApJ...789...48O}.  
We nonetheless note that the formal solution~\eqref{eq_step_open_SolAna} 
    has not been derived this way to our knowledge.      
We also note that this solution is valid at arbitrary $(x, t)$ 
    for arbitrary $k$.
}     

\subsection{Discrete Leaky Modes vs. Improper Continuum Eigenmodes}
\label{sec_sub_stepDLM}
This subsection aims at offering a better distinction between 
     discrete leaky modes (DLMs) and improper continuum eigenmodes. 

{ We start by summarizing some necessary properties of DLMs,
    which are solutions to an EVP of { the ``EVP open noIC'' type}. 
We define $\mui^2$ and $\mue^2$ in the same way as $\ki^2$ and $\ke^2$
   in Equation~\eqref{eq_step_defkikekappae}, 
   seeing $\omega$ as possibly complex-valued and
   assuming $-\pi/2 < \arg\mui, \arg\mue \le \pi/2$ \citep[e.g.,][]{2015ApJ...814...60Y}. 
Note that the time-dependence
   of any perturbation is written as $\Exp{-i\omega t}$
   in its Fourier decomposition.
The mode function in the exterior then writes
   $\Exp{i\mue x}$ to within a complex-valued factor,
   because the other independent solution $\Exp{-i\mue x}$ to Equation~\eqref{eq_EVP_schro} is connected with { incoming waves}.
All mode frequencies are discrete, obeying the 
   nominal DR \citep[e.g.,][]{2005A&A...441..371T,2015ApJ...814...60Y}
\begin{equation}
\label{eq_step_open_leakyDR}
i\mue = \mui \cot(\mui R).
\end{equation} 
Equation~\eqref{eq_step_open_leakyDR} is well known to allow
   solutions with purely imaginary $\mue$ when $k > k_{{\rm cutoff}, 1}$.
These solutions coincide with the proper eigensolutions to ``EVP open noBC'',
    and are therefore not of further interest.
More interesting are the DLMs, which exist regardless of $k$. 
As first shown by \citet{2005A&A...441..371T}, the complex-valued frequency 
    of the $m$-th DLM may be approximated by
\begin{equation}
\label{eq_step_DLM_omgmSmallK}
\begin{split}
&     \dfrac{R\Re\omega^{\rm DLM}_{m}}{\vai} 
 \approx
      \left(m-\dfrac{1}{2}\right)\pi, 
      \quad (m=1, 2, \cdots) \\
&     \dfrac{R\Im\omega^{\rm DLM}_{m}}{\vai} 
 \approx 
     -\dfrac{1}{2}\ln\dfrac{1+\sqrt{\rhoe/\rhoi}}{1-\sqrt{\rhoe/\rhoi}}, 
\end{split}
\end{equation}    
    provided that $|\omega^{\rm DLM}_m|^2 \gg k^2 \vae^2$.
That DLMs do not qualify as eigenmodes have been pointed out
    in the cylindrical study by \citetalias{2015ApJ...806...56O}.
The same holds for the Cartesian version, to corroborate which 
    we add that no scalar product can be properly
    defined for the relevant EVP.
That said, it is known that the periods and damping times predicted
    with Equation~\eqref{eq_step_open_leakyDR} may describe a substantial portion
    of the time sequences of the relevant perturbations
    (e.g., \citealt{2005A&A...441..371T,2014A&A...567A..24H,2015ApJ...814...60Y}).
Whether this happens is solely determined by how the initial perturbation $u(x)$
    is prescribed, a notion highlighted by \citetalias{2007PhPl...14e2101A}.
We will visualize this notion by several specific computations, which in a sense
    makes the notion more concrete.}       

\begin{figure}
\centering
 \includegraphics[width=.95\columnwidth]{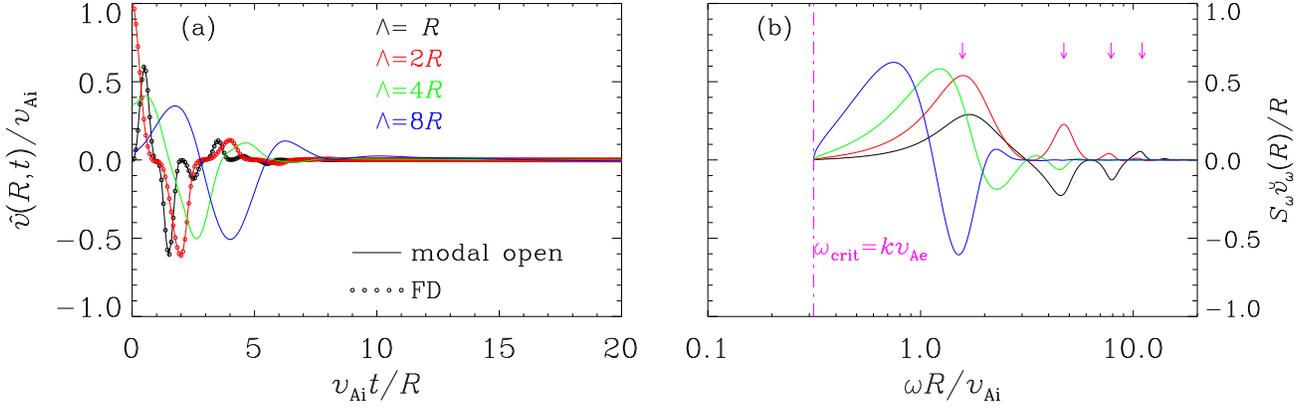}
 \caption{
 ``Modal Open'' solutions to IVP~\ref{ivp_mu_open} for a coronal slab with
     step density profile ($\mu=\infty$) and
     a density contrast of $\rhoi/\rhoe=2.25$.
 The dimensionless axial wavenumber is fixed at $kR = \pi/15$.               
 By ``Modal Open'' we mean that the solution is constructed by superposing
     the eigenmodes for an open system. 
 A number of values are examined for the spatial extent ($\Lambda$)
     of the initial perturbation as discriminated by the
     different colors.
 Plotted in (a) are the temporal profiles of the transverse speed
     at the nominal slab half-width, namely $\hat{v}(R, t)$.
 In addition to the ``Modal Open'' solutions (the solid curves), 
     the finite-difference solutions are also presented for two values
     of $\Lambda$ for comparison (the black and red symbols). 
 Presented in (b) are the frequency dependencies of the 
     local spectral densities at the nominal slab half-width, namely $S_\omega \breve{v}_\omega(R)$.           
 The vertical dash-dotted line represents the critical frequency
     $\omgcrit = k \vae$.
 The real parts of the mode frequencies for the first several
     discrete leaky modes (DLMs) are indicated by the magenta arrows.
 These DLMs are computed with Equation~\eqref{eq_step_open_leakyDR}, 
     the dispersion relation (DR) pertaining to an eigenvalue problem (EVP)
     that does not allow { incoming waves} at infity.
 The continuous curves, however, derive from the improper continuum that
     results from an EVP for which no boundary condition is specified at
     infinity.            
 See text for more details.
 }
 \label{fig_step_modal}
\end{figure}

Figure~\ref{fig_step_modal} presents the ``Modal Open'' solutions
    (Equation~\eqref{eq_step_open_SolAna}, the solid curves) to
    the $\mu=\infty$ version of IVP~\ref{ivp_mu_open}.
Only the improper continuum needs to be considered, given that
    $[\rhoi/\rhoe, kR]$ is fixed at $[2.25, \pi/15]$.
A number of values are examined for the spatial extent
    of the initial perturbation ($\Lambda$)
    as discriminated by the different colors. 
The lateral speeds at the nominal 
    half-width $\hat{v}(R, t)$ as functions of time
    are plotted in Figure~\ref{fig_step_modal}a, where the finite difference (FD)
    solutions are shown by the symbols for comparison. 
Note that the FD solutions are shown only for
    $\Lambda/R=1$ and $\Lambda/R=2$
    to avoid further crowding the curves.   
Nonetheless, it holds in general that the FD solutions agree exactly
    with the ``Modal Open'' ones, thereby validating both approaches. 
More importantly, one sees that the temporal profile involves increasingly
    fine scales as $\Lambda$ decreases.
This behavior is then quantified in Figure~\ref{fig_step_modal}b, where
    the local spectral densities $S_\omega \breve{v}_\omega(R)$ 
    are shown against the frequency $\omega$. 
The critical frequency $\omgcrit = k \vae$ 
    is given by the vertical dash-dotted line for reference.
In agreement with Figure~\ref{fig_step_modal}a,
    one sees that improper eigenmodes with increasingly high frequencies are involved
    when $\Lambda$ decreases. 
Furthermore, one sees that the local spectral densities tend to favor
    a set of discrete frequencies indicated by the arrows, which represent
    the real parts of the mode frequencies ($\{\Re\omega^{\rm DLM}_{m}\}$) 
    of the first several DLMs
    as found by numerically solving Equation~\eqref{eq_step_open_leakyDR}.
Some further inspection indicates that the values
    for the set $\{\Re\omega^{\rm DLM}_{m}\}$
    can be safely approximated
    by Equation~\eqref{eq_step_DLM_omgmSmallK}, which is not
    surprising given that the inequality $|\omega^{\rm DLM}_{m}|^2 \gg k^2 \vae^2$
    holds well. 
Equally unsurprising is that these frequencies tend to stand out
    in $S_\omega\breve{v}_\omega(R)$, for they tend to do so
    in $S_\omega$.
We capitalize on the expression for $S_\omega$
    (see Equation~\eqref{eq_step_open_Sol_DefcjSomg}) to explain this,
    and focus for now on the situation where $\omega^2 \gg k^2 \vae^2$.
It then follows that $\ki^2 \approx \omega^2/\vai^2$ and
    $\ke^2 \approx \omega^2/\vae^2$.
Consequently, the $\omega$-dependence of $S_\omega$ is largely determined
    by the factor $\Ac^2+\As^2$, which in turn approximates to
\begin{equation}
\Ac^2 +\As^2 \approx \rhoe/\rhoi + (1-\rhoe/\rhoi)\cos^2(\omega R/\vai).
\end{equation}        
Evidently, the set $\{\Re\omega^{\rm DLM}_{m}\approx (m-1/2)\pi \vai/R\}$ 
    minimizes $\Ac^2+\As^2$ and therefore tends to maximize $|S_\omega|$. 
That $S_\omega \breve{v}_\omega(R)$ tends to attain a local extreme at 
    the set $\{\Re\omega^{\rm DLM}_{m}\}$ is further strengthened
    by the fact that $\breve{v}_\omega(R)$ does so given
    that $\breve{v}_\omega(R)\approx -\vai\sqrt{\rhoe/\rhoi}\sin(\omega R/\vai)$ 
    (see Equation~\eqref{eq_step_open_vImproper}).
One therefore deduces with Equation~\eqref{eq_step_open_Sol_DefcjSomg} that
    the deviation of the locations of local extrema from 
    $\{\Re\omega^{\rm DLM}_{m}\}$ when $\omega^2\gg k^2 \vae^2$ 
    is almost entirely due to the specification of the initial perturbation. 
This deviation is reflected in two aspects.
Firstly, the positions of some extrema may be slightly different from
     $\{\Re\omega^{\rm DLM}_{m}\}$, with the first extremum for $\Lambda/R=1$
     or $\Lambda/R=2$ being an example. 
Secondly, and more severely, some extrema may not be accounted for by
     $\{\Re\omega^{\rm DLM}_{m}\}$.
This is particularly true when $\Lambda/R$ is large, as exemplified by 
     the second (third) extremum for $\Lambda/R = 4$ ($\Lambda/R = 8$).  
Now consider the situation where the inequality $\omega^2 \gg k^2 \vae^2$
     does not hold. 
{ For large $\Lambda$, one sees some further extrema that take place at
     frequencies below $\Re\omega^{\rm DLM}_{1}$.}
The first extremum for $\Lambda/R = 8$ serves as an example for this aspect, 
     and it therefore does not make much sense to further question
     whether the associated damping timescale agrees with the expectation with
     Equation~\eqref{eq_step_open_leakyDR}. 
Rather, with Figure~\ref{fig_step_modal} we conclude that the details of the initial
     perturbation play an important role in the temporal evolution 
     of the system.

\subsection{Connection Between the ``Modal Closed'' and ``Modal Open" Solutions}
\label{sec_sub_step_converge2cont}
This subsection discusses the connection between the ``Modal Closed''
    and ``Modal Open'' solutions, the aim being to further visualize 
    the fact that improper continuum modes qualify as eigenmodes. 
For simplicity, we focus on the situation where $k<k_{{\rm cutoff},1}$
    such that proper modes are irrelevant.    

Let us start with some analytical progress for the eigensolutions
    to EVP~\ref{evp_mu_closed}, which we recall is defined
    on a closed domain $[0, d]$.
The eigenfunctions remain expressible by
    Equation~\eqref{eq_step_open_vImproper}, except that the
    eigenfrequencies are discrete as determined by the DR
\begin{equation}
\label{eq_step_modalclosed_DR}
\Ac \cos(\ke d) + \As \sin(\ke d) = 0. 
\end{equation}    
Evidently, this DR is nothing but the BC at the outer boundary. 
Consider now low-frequency eigenmodes, by which we mean $\omega = k\vae(1+\Delta)$
    where $0<\Delta \ll 1$. 
One readily recognizes from Equation~\eqref{eq_step_defkikekappae} that 
    $\ki^2\approx k^2(\rhoi/\rhoe-1)$ and $\ke^2 \approx k^2 (2\Delta)$.
Now suppose $\ke R \ll 1$.
Equation~\eqref{eq_step_open_ACAS} then indicates that
    $\Ac/\As \sim \bigO(\ke R)$, meaning that the leading order solution
    to Equation~\eqref{eq_step_modalclosed_DR} is $\ke d \approx l\pi$.
Hence $\Delta \approx (l\pi)^2/(2 k^2 d^2)$.
To sum up, the low-frequency eigenmodes are characterized by 
\begin{equation}
\label{eq_step_eigenLow}
\dfrac{\omgl}{k\vae} \approx 1+\dfrac{(l\pi)^2}{2 (kd)^2}, 
\quad \text{provided that} \quad  
	(kd)^2 \gg (l\pi)^2, 
	\quad 
	d/R \gg l\pi.
\quad 
\end{equation}
Now consider high-frequency eigenmodes, by which we mean
    $\omega^2 \gg k^2 \vae^2$.           
Equation~\eqref{eq_step_defkikekappae} indicates that 
    $\ki \approx \ke \approx \omega/\vai$.     
Some algebra then leads the DR \eqref{eq_step_modalclosed_DR} to become
\begin{eqnarray}
\label{eq_step_DRHigh}
\sin\left[\ke(d-R)+\phi_{\rm e} \right] \approx 0,
\end{eqnarray}    
    where $\phi_{\rm e}$ is governed by
\begin{equation}
\tan\phi_{\rm e} = \sqrt{\rhoe/\rhoi}\tan(\sqrt{\rhoi/\rhoe}\ke R).
\end{equation}    
One readily sees that the leading order solution is given by 
    Equation~\eqref{eq_omglhigh}, which is not surprising because
    this approximate expression applies to arbitrary $\mu$.
Nonetheless, the explicit DR \eqref{eq_step_modalclosed_DR} can be employed to
    yield the next order correction, the net result being
\begin{equation}
\label{eq_step_omgHighcorr}
\begin{split}
& \omgl  \approx (l\pi - \phi_l) \dfrac{\vae}{d-R}, \\
& \phi_l \approx 
 	     \arctan\left[
         \sqrt{\rhoe/\rhoi}
            \tan
         	\left(\sqrt{\rhoi/\rhoe}\dfrac{l\pi}{d/R-1}\right)
         \right]
         +\left\lfloor\sqrt{\rhoi/\rhoe}\dfrac{l}{d/R-1}
                +\dfrac{1}{2}
          \right\rfloor 
          \pi, 
\end{split}
\end{equation}    
    with $\left\lfloor\cdot\right\rfloor$ being the floor function.
Equation~\eqref{eq_step_omgHighcorr} is presented only for completeness,
    the derivation for which is the same as
    in Appendix~A.2 of \citetalias{2022ApJ...928...33L}.
    
More important to this study is that 
    { the discrete set of eigenfrequencies ${\omgl}$}
    for EVP~\ref{evp_mu_closed} converges to the improper continuum
    when the domain size $d$ extends to infinity. 
Let 
\begin{equation}
\label{eq_def_freqSpacing}
\Delta\omgl = \omega_{l+1}-\omega_{l},
\quad 
l=1, 2, \cdots
\end{equation}
    define the frequency spacing. 
The continuum is then approached in such a way that $\Delta\omgl$ starts with
    a $1/d^2$ dependence 
    for small $l$ (see Equation~\eqref{eq_step_eigenLow}) and transitions to 
    a $1/d$ dependence for large $l$ (see Equation~\eqref{eq_omglhigh}).
{ In view of Equation~\eqref{eq_modal_formalSol},
    we define
\begin{eqnarray}
 \label{eq_def_SpecDen}  	   
  S_l
= 
   \dfrac{c_l}{\omega_{l+1}-\omega_{l}}
\end{eqnarray}    
   such that $S_l \breve{v}_l(x)$ represents some discrete local spectral density.
Let us demonstrate that the discrete set $\{S_l\breve{v}_l(x)\}$ 
    approaches the continuous ${S_\omega}\breve{v}_\omega(x)$
    in Equation~\eqref{eq_step_open_SolAna}.} 

\begin{figure}
\centering
 \includegraphics[width=.95\columnwidth]{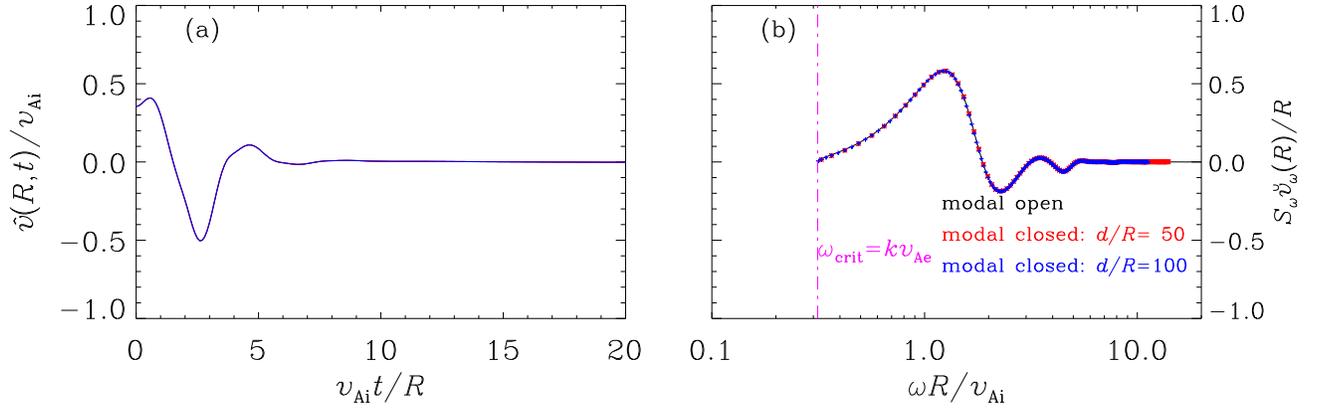}
 \caption{
 Similar to Figure~\ref{fig_step_modal} but for a fixed combination
     $[\rhoi/\rhoe, kR, \Lambda/R] = [2.25, \pi/15, 4]$. 
 The ``Modal Open'' solution is represented by the black curves in both
     (a) and (b).
 Additional ``Modal Closed'' solutions are presented as found by superposing
     the eigenmodes for the EVP defined on closed domains
     (EVP~\ref{evp_mu_closed}). 
 Two domain sizes are discriminated by the different colors, one
     being $d/R=50$ and the other being $d/R=100$.
 The eigenfrequencies for EVP~\ref{evp_mu_closed} are discrete by
     construction.    
 See text for more details.                  
 }
 \label{fig_step_cont}
\end{figure}

Figure~\ref{fig_step_cont} is similar in format to Figure~\ref{fig_step_modal},
    the primary difference being that the combination $[\rhoi/\rhoe, kR, \Lambda/R]$
    is now fixed at $[2.25, \pi/15, 4]$.
The ``Modal Open'' solution is shown by the black curves.
Two domain sizes are employed for constructing the ``Modal Closed'' solutions,
    one being $d/R=50$ (the curve and symbols in red),
    and the other being $d/R = 100$ (blue).
Examine Figure~\ref{fig_step_cont}a first, from which one sees that the ``Modal Open''
    and ``Modal Closed'' solutions cannot be told apart.
We therefore conclude with Figure~\ref{fig_step_cont}a 
    and Figure~\ref{fig_step_modal}a that the ``FD'', ``Modal Closed'',
    and ``Modal Open'' approaches yield identical solutions to IVP~\ref{ivp_mu_open}.
{ Now move on to Figure~\ref{fig_step_cont}b}, where the symbols
    represent  the values of $\{S_l \breve{v}_l(R)\}$ as functions
    of the discrete eigenfrequencies $\{\omgl\}$ associated with
    the ``Modal Closed'' solutions.
One sees that the set $\{\omgl\}$
    for $d/R=100$ is finer than for $d/R=50$,
    a feature that holds in view of
    Equations~\eqref{eq_step_eigenLow} and \eqref{eq_omglhigh}.
More importantly, one sees that the sets $\{S_l \breve{v}_l(R)\}$ for both values
    of $d/R$ cannot be distinguished from the continuous
    $S_\omega \breve{v}_\omega(R)$ associated with the improper continuum. 
We therefore conclude that the continuous $S_\omega \breve{v}_\omega(R)$    
    can be adequately resolved with the discrete eigenmodes for a domain size
    as modest as $d/R=50$. 
{ This also visualizes the general behavior
    for $\{S_l \breve{v}_l(x)\}$ to converge 
    to $S_\omega \breve{v}_\omega(x)$ at arbitrary $x$.} 

\subsection{Observational Implications}
{ 
Some observational implications follow from our discussions
    on the connection between DLMs and the improper continuum.
For clarity, we restrict our discussions to those observed signals that attenuate
    temporally. 
Our Figure~\ref{fig_step_modal}a suggests that the quasi-periods 
    are consistently on the order of the transverse \Alf\ time $R/\vai$, provided 
    that the spatial extent of the initial perturbation differs not too much
    from the width of the wave host. 
This supports the customary practice that candidate sausage perturbations are
    identified primarily by their short quasi-periods
    \citep[e.g.,][]{2003A&A...412L...7N,2015A&A...574A..53K,2019NatCo..10.2276C}.
Furthermore, the formal solution (Equation~\eqref{eq_step_open_SolAna}) stresses that the
    temporal damping derives from the destructive interference of 
    the monochromatic components in the improper continuum. 
It then follows that a rich variety of damping envelopes may result, with
    the details of the initial perturbation being critical
    (see the expression for $S_\omega$ in Equation~\eqref{eq_step_open_Sol_DefcjSomg}).
This damping envelope may be similar to an exponential one during some appropriate
    interval, and hence agree with the expectations with DLMs
    (e.g., Figure~4 in \citealt{2005A&A...441..371T}).
Conversely, nominal DRs of DLMs can be put to seismological use when
    some exponential damping shows up, an example being the analysis by
    \citet{2007AstL...33..706K} of the flare pulsation originally reported
    by \citet{1973SoPh...32..485M}. 
However, it is equally possible, if not more common, that the damping envelope
    follows different patterns, given the diversity of the initial perturbations
    (see Figure~\ref{fig_step_modal}a).
Two consequences then arise.
Firstly, there is no need to reject short-period signals as
    candidate sausage perturbations simply because their damping envelopes deviate
    from an exponential one.
It is just that the meaning of the deduced damping time needs to be clarified.
In this sense, one may need to re-assess the seismology performed by, say,
    \citet{2015ApJ...812...22C} for the pulsating flare emission
    presented in \citet{2015A&A...574A..53K}.
Our point is, the seismology therein associated the damping signal with
    a DLM.
The observed temporal damping, however, does not show a particular pattern 
    \citep[][Figure~3]{2015A&A...574A..53K}.
Secondly, the details of an observed time-varying signal actually encapsulate 
    the information on not only the wave host but also the initial perturbation. 
It is therefore possible, at least in principle, to decipher some appropriate
    flare QPPs to deduce some parameters that characterize the energy-release process.
The same thinking applies to AR loops as well, if a candidate sausage
    perturbation is eventually identified.     
}

\section{Continuous Density Profiles ($\mu < \infty$)}
\label{sec_finitemu}
This section examines the situation where $\mu$ is finite. 
We largely rely on the ``Modal Closed'' approach, and present
    the FD solutions only when necessary.
         
\subsection{Comparison Between the Cases with $\mu<2$ and $\mu>2$}
\label{sec_sub_finitemu_overview}
\begin{figure}
\centering
 \includegraphics[width=.7\columnwidth]{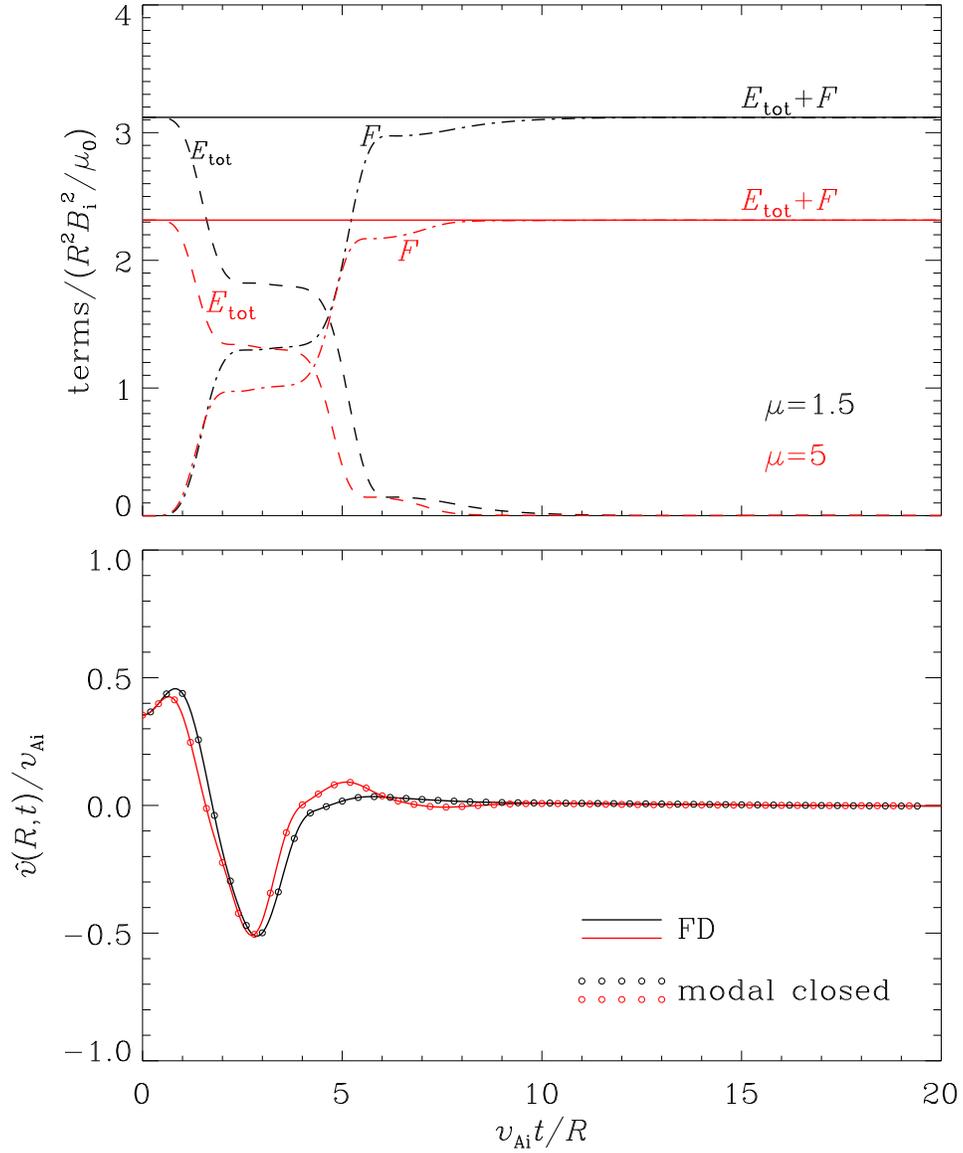}
 \caption{
 FD solutions to IVP~\ref{ivp_mu_open} for two steepness parameters,
     one being $\mu=1.5$ (the black curves)
     and the other being $\mu=5$ (red), 
     given a fixed combination $[\rhoi/\rhoe, kR, \Lambda/R] = [2.25, \pi/15, 4]$.
 Plotted in (a) are the terms characterizing the energy content 
     in the volume $\mathcal{V}$ where the initial perturbation is applied,
     with $E_{\rm tot}$ and $F$ representing the total wave energy in
     and the cumulative energy loss from $\mathcal{V}$
     (see Equations~\eqref{eq_ConsInt_Etot} to \eqref{eq_def_ener_F}).
 The transverse speeds at the nominal half-width $\hat{v}(R, t)$ 
     are shown against time in (b), where
     the ``Modal Closed'' solutions are presented by the symbols as labeled. 
 These solutions are obtained with Equation~\eqref{eq_modal_formalSol},
     with the domain size chosen to be $d = 50~R$.  
 }
 \label{fig_vtdep_FD_vs_modal}
\end{figure}

Figure~\ref{fig_vtdep_FD_vs_modal} presents the FD solutions
    to IVP~\ref{ivp_mu_open} for a fixed combination
    $[\rhoi/\rhoe, kR, \Lambda/R] = [2.25, \pi/15, 4]$.
Two steepness parameters, $\mu=1.5$ and $\mu=5$, are examined
    as discriminated by the different colors. 
Note that FD solutions are shown here to avoid discussing 
    the outer boundary, which is inherent to the ``Modal Closed'' approach. 
Figure~\ref{fig_vtdep_FD_vs_modal} offers some gross properties of the solutions,
    where by gross we choose to examine the energetics in
    the volume $\mathcal{V}$ within which the initial perturbation 
    is implemented. 
Specifically, the total energy $E_{\rm tot}$ in $\mathcal{V}$ and  
    the cumulative energy loss $F$ from $\mathcal{V}$
    are evaluated with Equations~\eqref{eq_def_ener_Etot}
    and \eqref{eq_def_ener_F} by letting $x=\Lambda$ therein.
One sees no qualitative difference between the solutions for different values
    of $\mu$, despite that these values lie astride the
    nominally critical value of two. 
In both cases, $E_{\rm tot}$ features some rather rapid decrease
    with time, accompanied by some steady increase in $F$.
The sum of the two quantities ($E_{\rm tot}+F$) remains constant throughout,
    indicating that energy is conserved remarkably well
    (see Equation~\eqref{eq_ConsInt_Etot}). 
In this regard, the plateaus in $E_{\rm tot}$ or $F$ 
    arise simply because the transverse component of the
    Poynting vector is small during the corresponding intervals
    (see Equation~\eqref{eq_def_ener_F}).
More importantly, $E_{\rm tot}$ drops to extremely small values when, say,
    $t\gtrsim 10R/\vai$ for both $\mu$. 
One then expects that the signals in the entire interval $x \le \Lambda$ 
    behave in a similar manner.
This is indeed true, as evidenced by Figure~\ref{fig_vtdep_FD_vs_modal}b 
    where we arbitrarily single out the transverse speed
    at the nominal slab half-width $\hat{v}(R, t)$.
One then sees again the rapid attenuation for both steepness parameters,
    with only two (four) extrema discernible when $\mu=1.5$ ($\mu=5$).
We remark that the symbols represent the ``Modal Closed'' solutions 
    constructed with Equation~\eqref{eq_modal_formalSol} with the domain size
    chosen to be $d/R = 50$. 
That the FD and ``Modal Closed'' solutions agree exactly means that 
    this modest domain size is adequate for describing the behavior
    of $\hat{v}(R, t)$ in the examined timeframe. 
        
\begin{figure}
\centering
 \includegraphics[width=.7\columnwidth]{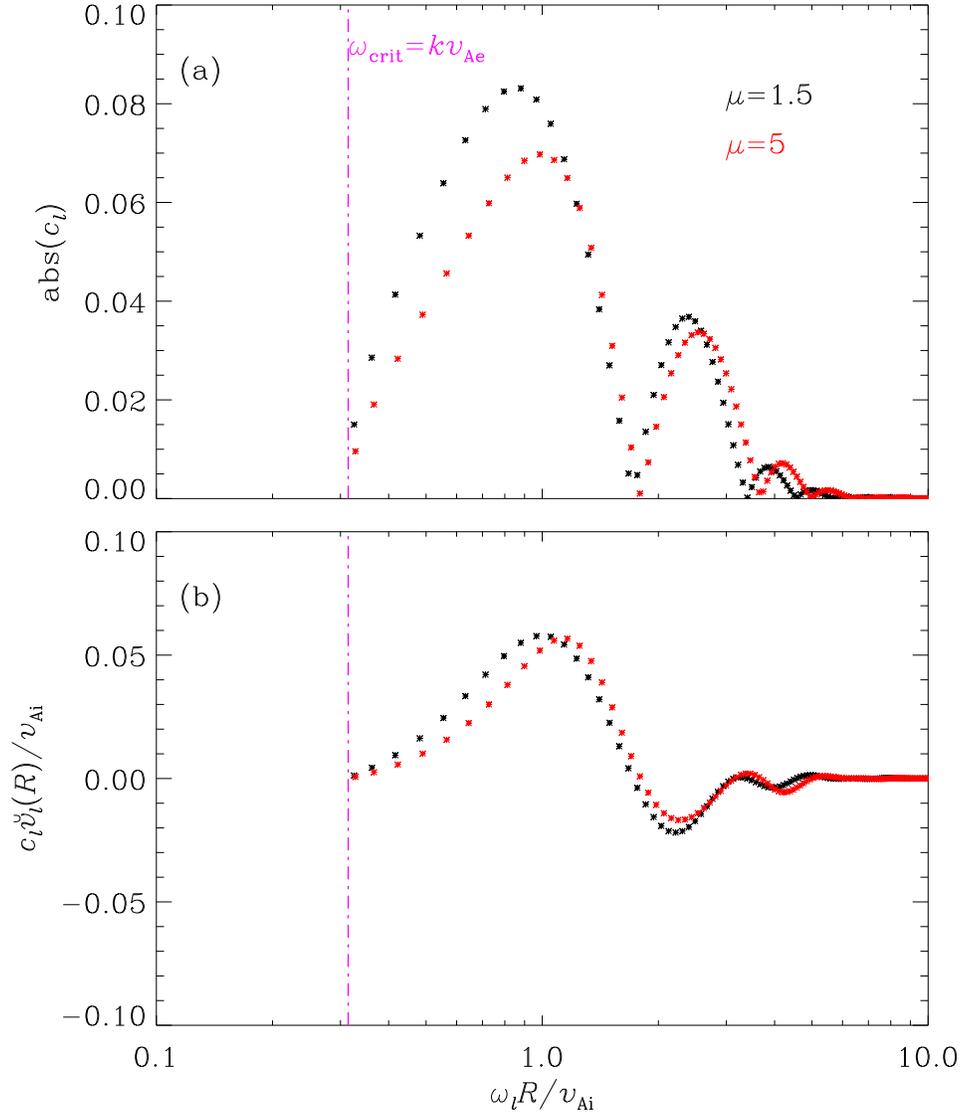}
 \caption{
Frequency-dependencies of 
    the contributions of individual modes 
    for a fixed combination $[\rhoi/\rhoe, kR, \Lambda/R, d/R] = [2.25, \pi/15, 4, 50]$.
 Two steepness parameters, $\mu=1.5$ and $\mu=5$, are examined as labeled.
 The sets of position-independent coefficients $\{c_l\}$ are plotted in (a),
     while the specific contributions at the half-width $c_l \breve{v}_l (R)$ 
     are presented in (b).      
 The critical frequency $\omgcrit = k \vae$ is represented 
    by the vertical dash-dotted lines for reference.
 }
 \label{fig_ME_modal_contrib}
\end{figure}

Figure~\ref{fig_ME_modal_contrib} capitalizes on 
    Equation~\eqref{eq_modal_formalSol} to examine the contributions
    from individual eigenmodes.
For this purpose, the discrete sets of position-independent coefficients $\{c_l\}$
    and the specific contributions at the half-width $\{c_l \breve{v}_l(R)\}$    
	are grabbed from the ``Modal Closed'' solutions from
	Figure~\ref{fig_vtdep_FD_vs_modal}.
They are then plotted against the sets of eigenfrequencies $\{\omgl\}$
    in Figures~\ref{fig_ME_modal_contrib}a 
    and \ref{fig_ME_modal_contrib}b, with the two different steepness parameters
    again discriminated by the different colors. 
The critical frequency $\omgcrit = k\vae$ is given by the vertical dash-dotted
    lines for comparison. 
Examine Figure~\ref{fig_ME_modal_contrib}a first.
One sees that the frequency-dependency of $|c_l|$ for $\mu=1.5$ 
    is qualitatively the same as that for $\mu=5$, both featuring a series of peaks 
    at frequencies substantially higher than $\omgcrit$.
Now examining Figure~\ref{fig_ME_modal_contrib}b, one sees 
    the same set of peak frequencies in $\{c_l \breve{v}_l(R)\}$ 
    for each examined $\mu$.
The values of these peak frequencies then explain why the transverse \Alf\ 
    time $R/\vai$ characterizes the periodicities in the signals themselves
    ($\hat{v}(R, t)$) shown in Figure~\ref{fig_vtdep_FD_vs_modal}b.
It is just that the term ``periodicity'' applies only marginally 
    to $\hat{v}(R, t)$ for $\mu=1.5$ to say the best, given the extremely rapid
    attenuation of the signal. 
Regardless, more important is that all eigenfrequencies are above $\omgcrit$, meaning
    that all eigenmodes are oscillatory. 
This is true not just for $\mu=5$ but for $\mu=1.5$ at least
    at the chosen domain size for EVP~\ref{evp_mu_closed},  
    despite that evanescent eigenmodes are ensured for an infinite domain 
    when $\mu <2$.

\subsection{Dependence of Modal Structure on Domain Size}
\label{sec_sub_finitemu_modalStr}
This subsection examines how the modal structure depends on the domain size $d$
    for EVP~\ref{evp_mu_closed}, where by ``modal structure'' we refer to
    the $l$-dependence of the eigenfrequency $\omgl$. 
Our purposes are twofold.
Firstly, we will explore how the solutions to EVP~\ref{evp_mu_closed}
    connect to the theoretical results on cutoff wavenumbers
    for an open system by \citetalias{2015ApJ...801...23L}.
In particular, we will visualize what is meant by the irrelevance
    of cutoff wavenumbers when $\mu<2$, paying special attention to
    how the number of evanescent eigenmodes varies with $d$ to evaluate
    the assertion that ``all fast modes are trapped''.
Secondly, in physical terms we will demonstrate when
    evanescent eigenmodes are guaranteed to be negligible as far as 
    the time-dependent solutions to IVP~\ref{ivp_mu_open}
    are concerned. 

\begin{figure}
\centering
 \includegraphics[width=.7\columnwidth]{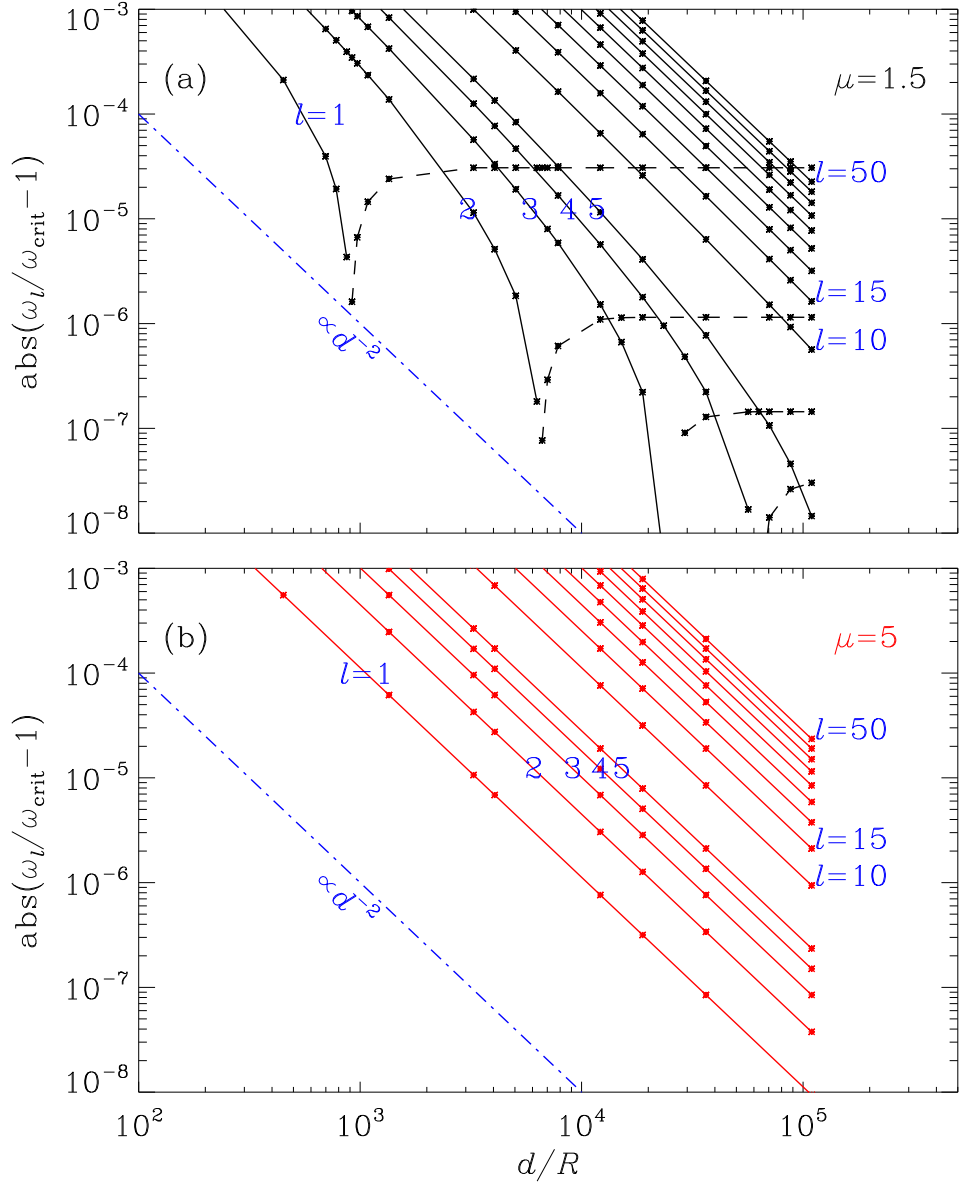}
 \caption{
Dependencies of the   	
   modulus of the fractional difference  
   $\delta_l = \omega_l/\omgcrit-1$ 
   on the dimensionless domain size ($d/R$)
   for two steepness parameters, 
   (a) $\mu = 1.5$ and (b) $\mu = 5$,
   given a combination $[\rhoi/\rhoe, kR] = [2.25, \pi/15]$. 
For each pair $[\mu, d/R]$, the first five eigenmodes are always presented, while
   the rest are uniformly sampled in $l$ with a step of five when $l$
   lies in the range between $10$ and $50$. 
A solid (dashed) curve is employed to connect $|\delta_l|$ for a given $l$
   when $\delta_l$ is positive (negative).
The blue dash-dotted lines represent the $1/d^2$-dependence for comparison.
See text for more details.      
}
 \label{fig_modStr_fracDiff}
\end{figure}

\begin{figure}
\centering
 \includegraphics[width=.95\columnwidth]{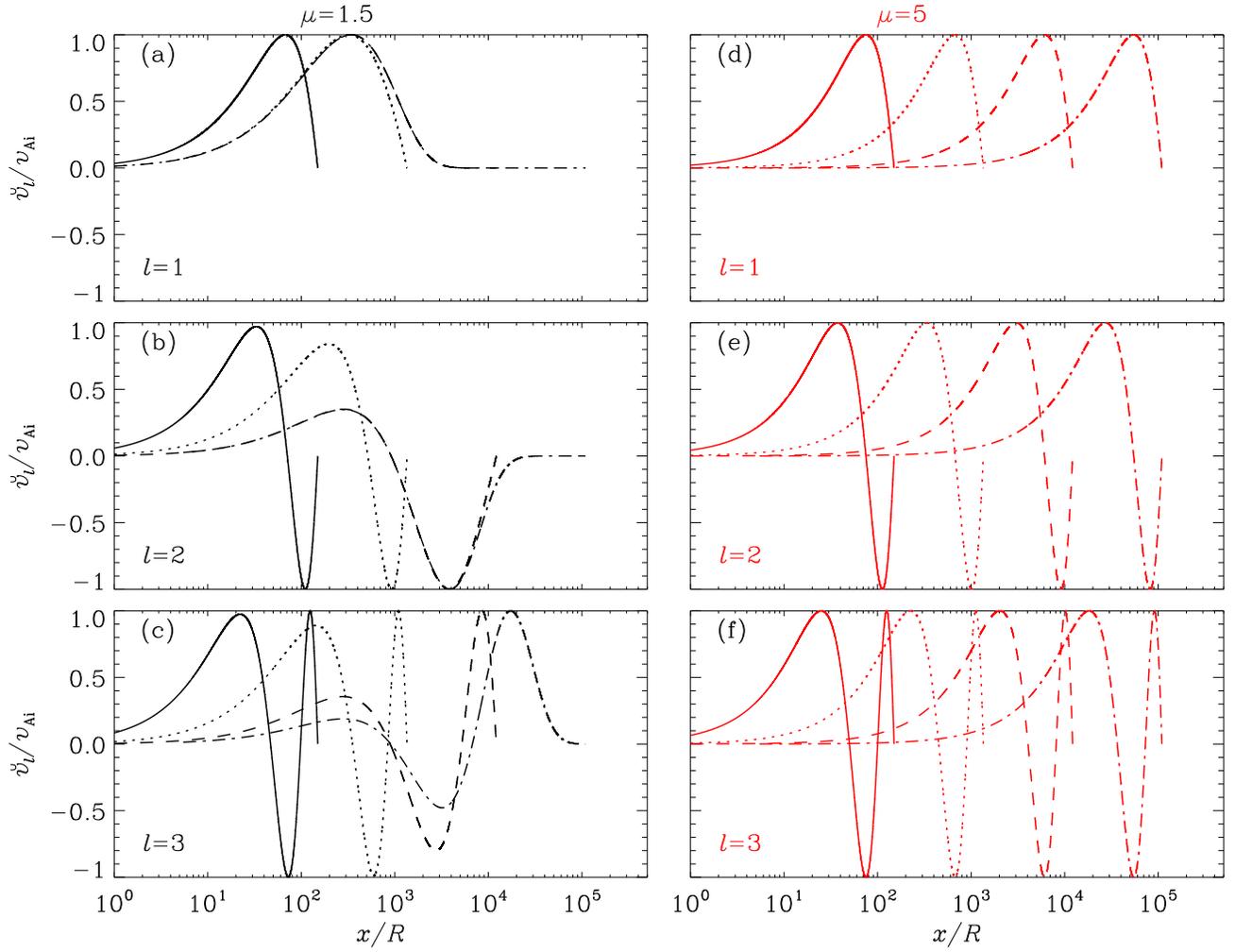}
 \caption{
 First three eigenfunctions for EVP~\ref{evp_mu_closed}
    on a number of domains differentiated 
    by the linestyles, given a 
    fixed combination $[\rhoi/\rhoe, kR]=[2.25, \pi/15]$.
 Two steepness parameters are examined, 
              one being $\mu = 1.5$ (the left column)
    and the other being $\mu = 5$ (right).
 The eigenfunctions are arbitrarily scaled such that their dependencies
    on the domain size can be better visualized.
 See text for more details.     
 }
 \label{fig_modStr_eigFunc}
\end{figure}

Defining some fractional difference $\delta_l = \omgl/\omgcrit-1$, 
    Figure~\ref{fig_modStr_fracDiff} examines how its modulus varies in response 
    to the variation of the dimensionless domain size $d/R$. 
The combination $[\rhoi/\rhoe, kR]$ is fixed at $[2.25, \pi/15]$, 
    whereas two different values are examined for the steepness parameter, 
    one being $\mu=1.5$ (Figure~\ref{fig_modStr_fracDiff}a)
    and the other being $\mu=5$ (Figure~\ref{fig_modStr_fracDiff}b). 
For each $\mu$, we choose to always present the first five eigenmodes but 
    sample other eigenmodes with a uniform step of five in $l$ when the mode number
    $l$ ranges from $10$ to $50$.  
The values for $\delta_l$ for any given $l$ are connected by a solid (dashed) curve
    when $\delta_l$ is positive (negative).     
The dash-dotted curves, on the other hand, represent a $1/d^2$ dependence
    as inspired by the analytical result for step density profiles 
    (see Equation~\eqref{eq_step_eigenLow}). 
Let us start with the case $\mu=5$ as presented in Figure~\ref{fig_modStr_fracDiff}b, 
    where one sees the unsurprising result that $\delta_l$ is consistently positive.
All eigenmodes are therefore oscillatory,
    a behavior that can be better visualized by the spatial profiles of
    the first three eigenfunctions shown in 
    the right column of Figure~\ref{fig_modStr_eigFunc}.   
It is also unsurprising to see that $\delta_l$ decreases monotonically
    with the domain size.
What Figure~\ref{fig_modStr_fracDiff}b demonstrates for sufficiently large $d/R$
    is that the frequency spacing $\Delta\omega_{l}$ starts with 
    a $1/d^2$ dependence for small $l$ 
    and gradually transitions to a $1/d$ dependence for large $l$, 
    with the latter dependence following from Equation~\eqref{eq_omglhigh}.
This behavior is qualitatively the same as happens for the step density profiles
    (see Section~\ref{sec_sub_step_converge2cont}).
We therefore conclude that a genuine continuum $(\omgcrit, \infty)$
    can be properly defined
    for an open system, provided that the limit $d/R\to\infty$ is taken.    
However, the modal behavior for $\mu=1.5$ is qualitatively different.
One sees from Figure~\ref{fig_modStr_fracDiff}a that evanescent eigenmodes
    indeed show up eventually for large enough $d/R$, as indicated by
    the changes of sign of $\delta_l$.
For instance, the first eigenmode transitions from an oscillatory to an evanescent one
    at $d/R \approx 900$, and so does the second (third) eigenmode at
    $d/R \approx 6600$ ($d/R \approx 29000$).
Once again, this behavior is made clearer by the spatial dependencies
    of the first several eigenfunctions (see the left column of Figure~\ref{fig_modStr_eigFunc}).      
One then deduces that an increasing number of evanescent eigenmodes
    emerge from the sea of
    oscillatory eigenmodes as the domain size increases, which we take as
    a visualization of the notion that cutoff wavenumbers are indeed irrelevant
    when $\mu<2$.
However, this notion needs to be understood as meaning that evanescent eigenmodes
    are always permitted for sufficiently large domains, regardless of
    the combination $[\rhoi/\rhoe, kR]$.
On the other hand, with the exception of the lowest subset,     
    the frequencies of oscillatory eigenmodes are such that
    the frequency spacing $\Delta\omgl$ also follows a $1/d^2$ dependence at small $l$
    and gradually gives way to a $1/d$ dependence at large $l$.
Despite this $d$-dependence, we conclude that a genuine continuum 
    $(\omgcrit, \infty)$
    cannot be defined, at least
    not by the straightforward limiting procedure $d/R\to\infty$ because
    the critical frequency is continuously crossed when $d/R$ increases
    continuously.  
We also argue that it does not much sense to explore the spectrum 
    for a truly open system per se, because all necessary physics 
    can be addressed by the ``Modal Closed'' approach, which is simple both
    technically and conceptually. 
In particular, one safely concludes that evanescent eigenmodes makes no contribution
    to the solution to IVP~\ref{ivp_mu_open} in Figure~\ref{fig_vtdep_FD_vs_modal}
    as far as the timeframe therein is of interest, despite that a large number 
    of evanescent eigenmodes are permitted for extraordinarily large domains. 
This conclusion holds because the solution in the pertinent timeframe is entirely
    contributed by oscillatory eigenmodes by any practical standard, 
    and evanescent eigenmodes are incomplete
    no matter how dense the set of their frequencies may be. 

\begin{figure}
\centering
 \includegraphics[width=.7\columnwidth]{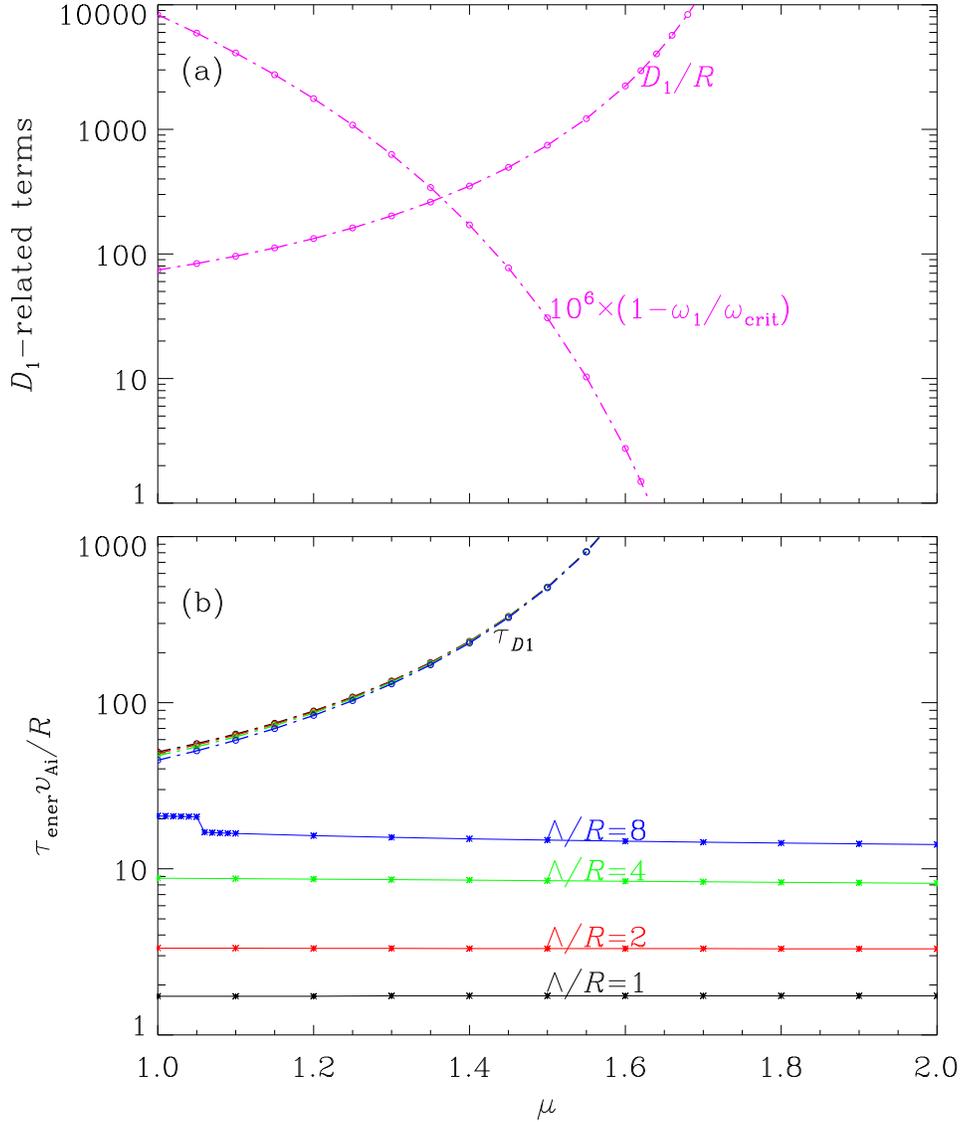}
 \caption{
Some scales characterizing the solutions to EVP~\ref{evp_mu_closed}
    and IVP~\ref{ivp_mu_open} for a fixed
    pair $[\rhoi/\rhoe, kR] = [2.25, \pi/15]$.      
Plotted in (a) is the $\mu$-dependence of $D_1$, the distance beyond which
    the first eigenfunction is evanescent.       
The fractional difference
    $1-\omega_1/\omgcrit$ is also displayed for reference. 
Panel (b) presents, by the solid curves, 
    the variations of some timescale $\tau_{\rm ener}$
    against $\mu$ for a number of $\Lambda$ as determined
    from the FD solutions to IVP~\ref{ivp_mu_open}.  
Here $\tau_{\rm ener}$ denotes the time at which 
    the total energy drops by a factor of $\Exp{4}\approx 55$ relative to
    its initial value in the volume where the initial perturbation is applied.
The dash-dotted curves represent the $\mu$-dependencies of
    another $\Lambda$-related timescale $\tau_{D1}$, 
    the time when the position $x=D_1$ is disturbed.    
The curves for both $\tau_{\rm ener}$ and $\tau_{D1}$ are 
    color-coded by the values of $\Lambda$.
See text for more details.    
 }
 \label{fig_D1related}
\end{figure}

We further exploit the ``Modal Closed'' approach to examine the condition
    for evanescent eigenmodes to be negligible in terms of their contributions
    to the system evolution.
This discussion turns out to be not that straightforward.
We therefore decide
    to employ a fixed combination $[\rhoi/\rhoe, kR] = [2.25, \pi/15]$
    and additionally assume $\Lambda\ge R$.
Only the situation $\mu<2$ is relevant now. 
Consider some extremely large domain size $d$ such that
    a substantial number of evanescent eigenmodes solve EVP~\ref{evp_mu_closed}.
Let the $l$-th evanescent eigenmode be characterized by some spatial scale $D_l$,
    which can be unambiguously identified as the position
    beyond which the $l$-th eigenfunction is evanescent.
It follows from Equation~\eqref{eq_def_QV} that 
\begin{equation}
\label{eq_def_Dl}
  \dfrac{D_l}{R} 
= \left[\dfrac{(1-\alpha_l)(\rhoi/\rhoe-1)}{\alpha_l}
      \right]^{1/\mu},
\end{equation}
    where $\alpha_l = 1-\omega_l^2/\omgcrit^2$. 
Evidently, $D_1 < D_2 <\cdots$ for a given $\mu$ 
    because $\omega_{1}<\omega_{2}<\cdots$. 
For definiteness, we evaluate $\omega_{1}$ at some $d$ beyond which
    $\omega_{1}$ does not vary with the domain size within the accuracy of our
    EVP solver (see Figure~\ref{fig_modStr_fracDiff}a).   
Figure~\ref{fig_D1related}a then displays $1-\omega_{1}/\omgcrit$
    and $D_1$ against $\mu$. 
One sees that $1-\omega_{1}/\omgcrit$ decreases
    and consequently $D_1$ increases
    extremely rapidly with $\mu$, the reason being that
    $\omega_{1}$ becomes increasingly close to $\omgcrit$ from below
    when $\mu$ approaches two.

Our following discussions center around $D_1$.      
Immediately ensuing is a timescale $\tau_{D1}$, namely the time at which the location
    $x=D_1$ is disturbed. 
Evidently,     
\begin{equation}
\label{eq_def_tauD1}
\tau_{D1} = \int_{\Lambda}^{D_1}\dfrac{dx}{\va(x)}.
\end{equation}   
Evanescent eigenmodes are guaranteed to play no role
    in the timeframe $t<\tau_{D1}$.
Figure~\ref{fig_D1related}b offers an example where we survey the 
    time-dependent solutions for a considerable number of combinations
    of $\mu$ and $\Lambda$.  
Note that the FD approach is adopted for convenience. 
For each $[\mu, \Lambda]$, we determine a further timescale 
    $\tau_{\rm ener}$ as the time at which
    $E_{\rm tot}(\Lambda, t)/E_{\rm tot}(\Lambda, t=0)$ 
    decreases to some threshold $\delta_{\rm ener}$ with
    $E_{\rm tot}(\Lambda, t)$ being the instantaneous total energy
    in the volume where the initial
    perturbation is applied (see Figure~\ref{fig_vtdep_FD_vs_modal}).
The solid curves represent the $\mu$-dependencies
    of $\tau_{\rm ener}$ determined with $\delta_{\rm ener}=\Exp{-4}\approx 1/55$
    for a number of $\Lambda$ as discriminated by the different colors. 
One sees that $\tau_{\rm ener}$ thus determined for a given $\Lambda$ is largely
    insensitive to $\mu$, the only exception being for $\Lambda/R=8$
    in the portion $\mu\lesssim 1.1$. 
There is actually some subtlety associated with this exception, as will be discussed
    shortly.
It suffices for now to notice the overall behavior 
    that $\tau_{\rm ener}$ increases with $\Lambda$ at a given $\mu$.
This behavior is understandable given that, when $\Lambda$ increases,
    the eigenmodes with increasingly low
    frequencies make more contributions to the system evolution, 
    thereby increasing the duration it takes for the eigenmodes to interfere to weaken
    $E_{\rm tot}(\Lambda, t)$ to some threshold. 
Regardless, one sees that $\tau_{\rm ener}$ at any examined $[\mu, \Lambda]$
    is lower than the corresponding $\tau_{D1}$, which is represented by 
    the dash-dotted curves and color-coded by $\Lambda$ in the same way
    as for $\tau_{\rm ener}$.
This is not surprising because $\tau_{\rm ener} \lesssim 20.8 R/\vai$
    with the maximum reached at $[\mu, \Lambda/R] = [1, 8]$, 
    while Equation~\eqref{eq_def_tauD1} indicates that 
    $\tau_{D1} > (D_1-\Lambda)/\vae$ and the latter attains
    a minimum of $\sim 44.2~R/\vai$ at the same $[\mu, \Lambda/R]$.
The comparison between $\tau_{\rm ener}$ and $\tau_{D1}$ 
    therefore validates the negligible role of evanescent eigenmodes, 
    which was explicitly shown by the specific ``Modal Closed'' solution for
    $[\mu, \Lambda/R] = [1.5, 4]$ 
    (see Figures~\ref{fig_vtdep_FD_vs_modal} and \ref{fig_ME_modal_contrib}).

We now come back to the behavior of $\tau_{\rm ener}$ 
    for $\Lambda/R=8$ as shown in Figure~\ref{fig_D1related}b.
By subtlety we refer to the apparent discontinuity at $\mu\approx 1.1$, 
    and refer further to the behavior of signals for $x\le\Lambda$
    in the interval $t < \tau_{D1}$.
We restrict ourselves to the parametric survey presented in
    Figure~\ref{fig_D1related} for the ease of description. 
The signals turn out to always evolve in three stages, to be labeled I to III hereafter.
Stage I is characterized by some rapid attenuation, and the periodicity
    tends to be substantially shorter than $P_{\rm longi} = 2\pi/\omgcrit$. 
Here the subscript ``longi'' is meant to indicate that this periodicity
    evaluates to the longitudinal \Alf\ time $2L/\vae$ for axial fundamentals.    
In addition, $E_{\rm tot}(\Lambda, t)$ hardly shows any increase with time.    
Stage III, on the other hand, is characterized by a nearly
    monochromatic variation with the periodicity
    being very close to $P_{\rm longi}$.
The signals still weaken with time but in a fashion different from Stage I,
    meaning in particular that $E_{\rm tot}(\Lambda, t)$ is wavy
    and its envelope declines at a much lower rate. 
Stage II corresponds to the transition from Stage I to Stage III, and 
    tends to be short in duration.
The appearance of the three stages is actually a natural result of
    Equation~\eqref{eq_modal_formalSol}.   
Stage~I tends to involve eigenmodes with frequencies $\omgl$ substantially higher
    than $\omgcrit$, and the signals therefore tend to attenuate rapidly because
    the involved monochromatic components can readily interfere to largely cancel
    out their contributions. 
Consequently, modes with frequencies very close to $\omgcrit$ are allowed to stand
    out in stage III, and the associated monochromatic components require longer time
    to interfere to yield some appreciable attenuation.
Note that the physics discussed herein only requires the negligible role
    of evanescent eigenmodes, and hence applies to the case for $\mu>2$ when
    $k<k_{{\rm cutoff},1}$.
It is just that evanescent eigenmodes and $\tau_{D1}$ become irrelevant altogether
    in that case, and there is no need to consider
    any complication beyond $t=\tau_{D1}$.  
Now focus once again on Figure~\ref{fig_D1related}b. 
The details of Stage II, in particular where it is positioned in time, 
    turn out to possess a rather intricate dependence on $\mu$ and $\Lambda/R$.
One end result is that $\tau_{\rm ener}$ always appears in Stage I
    for all examined pairs $[\mu, \Lambda/R]$ with the exception being
    for $\mu\lesssim 1.1$ when $\Lambda/R=8$.
In this case $\tau_{\rm ener}$ appears in Stage II, and the somehow
    wavy behavior in $E_{\rm tot}(\Lambda, t)/E_{\rm tot}(\Lambda, t=0)$
    means that the threshold $\delta_{\rm ener}$ is reached on the opposite side 
    of a ripple to what happens for larger $\mu$. 
Hence the discontinuity in the $\mu$-dependence of $\tau_{\rm ener}$.

\subsection{Further Remarks}
{ The results in this section can be further digested as follows.
We consistently see the interval $t\le \tau_{\rm ener}$ as the duration of interest,
   and restrict ourselves to 
   Figures~\ref{fig_vtdep_FD_vs_modal} and~\ref{fig_D1related}.  
Let us start with the FD solutions in Figure~\ref{fig_vtdep_FD_vs_modal} 
   and recall that no outer boundary is involved in IVP~\ref{ivp_mu_open}.
The drop of $E_{\rm tot}(t)/E_{\rm tot}(t=0)$ to the designated threshold
   $\delta_{\rm ener}$ can then be explained by that almost all of the wave energy
   in the volume $x\le \Lambda$ is carried away by the first two outgoing wave fronts. 
The passage of either front through $x=\Lambda$ is readily identified by 
   a rapid drop in $E_{\rm tot}(t)$ in front of a plateau-like feature.
This behavior happens for both $\mu=1.5$ and $\mu=5$, and holds also
   for the cylindrical computations (see Figure~2 in
   \citetalias{2022ApJ...928...33L}).
Now consider the ``Modal Closed'' solutions, and let $D_{\rm ener}$
   denote the distance that the outermost perturbation reaches
   when $t = \tau_{\rm ener}$.
Furthermore, let $D_{\rm ener}$ be chosen as the domain size
   for EVP~\ref{evp_mu_closed}.
By construction, the FD solution for $t\le \tau_{\rm ener}$ is
   identical to the ``Modal Closed''
   solution~\eqref{eq_modal_formalSol} constructed 
   with eigensolutions to EVP~\ref{evp_mu_closed}.   
Consequently, the outgoing fronts in the FD solution for $\mu=5$
   are in fact a manifestation of oscillatory eigensolutions, 
   which are the only solutions to EVP~\ref{evp_mu_closed} regardless of
   the domain size. 
Let us move on to $\mu=1.5$, for which Figure~\ref{fig_D1related}b shows that
   $\tau_{D1} \gg \tau_{\rm ener}$.
Hence $D_1 \gg D_{\rm ener}$.
The outgoing fronts in the FD solution for $\mu=1.5$ then also derive
   from the oscillatory eigensolutions to EVP~\ref{evp_mu_closed}, given that
   only eigensolutions of this type are allowed for any $d<D_1$. 
All in all, only oscillatory eigensolutions are allowed for both $\mu=1.5$ and $\mu=5$
   if $d=D_{\rm ener}$.
This then explains the qualitative similarity in the behavior
   of $E_{\rm tot}$ in the two FD solutions when $t<\tau_{\rm ener}$.
   
Some additional implications follow.
Our key message is that IVP studies in general offer a fuller picture
   for the wave dynamics than EVP analyses. 
Take the case $\mu=1.5$ for instance. 
Let $d' \gg D_1$ denote another domain size for EVP~\ref{evp_mu_closed}.
A considerable number of evanescent eigensolutions may be allowed together
    with an extremely closely spaced set of oscillatory ones.  
When $t<\tau_{\rm ener}$, however, the superposition of 
    all eigensolutions necessarily yields the same temporal behavior
    as computed with eigensolutions to EVP~\ref{evp_mu_closed}
    on the domain of size $d=D_{\rm ener}$.
In other words, evanescent eigenmodes can be safely
    discarded for both the gross energetics and 
    the signals sampled at individual locations
    (see Figure~\ref{fig_vtdep_FD_vs_modal}).
This then calls into question the seismological applications of the EVP
    analyses for $\mu<2$ by, say, \citet{2015ApJ...810...87L}.
Of interest therein was a pulsating flare event observed by the 
    Nobeyama Radioheliograph (NoRH) and originally reported in
    \citet{2013SoPh..284..559K}. 
Three periodicities were found, and the spatial distributions
    of individual power spectral densities favor their identification
    as the first three axial harmonics.
The interpretation of these signals as sausage perturbations was rejected
    by \citet{2013SoPh..284..559K}, for the measured
    periods are difficult to reconcile with sausage modes in flare loops with step transverse profiles. 
However, \citet{2015ApJ...810...87L} argued that this difficulty may be circumvented if
    one invokes transverse profiles similar to our ``outer $\mu$'' formulation with $\mu<2$.
We remark that evanescent eigensolutions were implied to dominate the
    system behavior for this argument to apply. 
Our results in this section indicate otherwise, 
    stressing that the physical relevance of a mode needs to be established
    by IVP studies before it is invoked for seismology. 
}

\section{Summary}
\label{sec_conc}

This study was motivated by some recent theoretical progresses
    that may cast doubt on the { well-known lack of candidate}
    sausage perturbations in solar active region (AR) loops. 
We adopted linear, 
    pressureless, ideal MHD to examine the responses of straight, field-aligned,
    axially uniform, coronal slabs to localized initial velocity perturbations 
    that excite axial fundamentals.  
The equilibrium density ($\rho_0$) was taken to be structured 
    only in one lateral direction ($x$), 
    following an ``outer $\mu$'' prescription where 
    { a steepness parameter $\mu$ pertains to the exterior (see Equation~\eqref{eq_rho_prof_outermu})}. 
An initial value problem (IVP, see IVP~\ref{ivp_mu_open}) 
    was formulated on the open domain $[0, \infty)$, to which 
    two types of eigenvalue problems (EVPs) are relevant. 
No boundary condition (BC) is specified at infinity in the ``EVP open noBC'' type,
    whereas the BC at infinity for { the ``EVP open noIC'' type 
    is that no incoming waves} are permitted. 
While working on an EVP of the ``EVP open noBC'' type, 
    \citet[][\citetalias{2015ApJ...801...23L}]{2015ApJ...801...23L}
    { showed that the steepness of $\rho_0$ in the exterior 
    is critical to the existence of cutoff axial wavenumbers.}
{ An application of the \citetalias{2015ApJ...801...23L} analysis
    to our ``outer $\mu$'' profile 
    yields that cutoff wavenumbers arise only when $\mu \ge 2$.
The lowest cutoff exceeds axial wavenumbers expected 
    for typical AR loops \citep{2018ApJ...855...53L}.}
Evanescent eigenmodes { (or equivalently trapped modes)
    are therefore prohibited, supporting the customary argument}
    for sausage perturbations to be difficult to detect.       
When $\mu<2$, however, cutoff wavenumbers are irrelevant in that 
    evanescent eigenmodes are always allowed regardless of
    the axial wavenumber ($k$) or the density contrast ($\rhoi/\rhoe$). 
{ This led \citetalias{2015ApJ...801...23L} to
    further argue that all fast waves are evanescent.
By implication, sausage perturbations may then} have been overlooked in AR loops with
    $\mu<2$.    
We focused on a value of $k$ that ensures the absence of
    evanescent eigenmodes for $\mu \ge 2$, and solved 
    IVP~\ref{ivp_mu_open} by superposing the eigenmodes on a closed domain 
    (EVP~\ref{evp_mu_closed}; also Equation~\eqref{eq_modal_formalSol}). 
Our study can be summarized as follows.

While revisiting IVP~\ref{ivp_mu_open} for step density profiles ($\mu\to\infty$),
    we offered an essentially analytical solution in terms of eigensolutions
    to the EVP of the ``EVP open noBC'' type. 
{ The importance of initial perturbations was demonstrated.
In particular, their specification impacts whether
    the system evolution can be understood with 
    discrete leaky modes, which arise for ``EVP open noIC''}.
Our ``Modal Closed'' solutions
    { then complemented previous studies of \citet{2007PhPl...14e2101A}
    and \citet{2015ApJ...806...56O}   
    by visualizing how the oscillatory continuum in ``EVP open noBC''
    is approached by the discrete eigenspectrum.} 
We further applied the ``Modal Closed'' approach to the case where $\mu$ is finite,
    and showed that the situation with $\mu \ge 2$ is qualitatively the same as in
    the step case.
Evanescent eigenmodes are prohibited by the relevance
    of cutoff axial wavenumbers, and an oscillatory continuum can result
    for open systems. 
When $\mu <2$, we showed that cutoff wavenumbers are indeed irrelevant. 
However, we stressed that this irrelevance does not 
    { mean that all fast waves are evanescent for truly open systems.
Rather, this means that an increasing number of evanescent
    eigenmodes result when the domain size increases for EVP~\ref{evp_mu_closed}. 
Oscillatory eigenmodes are always permitted, although} a genuine
    continuum cannot be obtained simply by letting the domain size approach infinity.
We quantified the condition for oscillatory eigenmodes to dominate
    the solutions to IVP~\ref{ivp_mu_open}, justifying the absence
    of any qualitative change of the system behavior when $\mu$ crosses
    the nominal critical value of two. 
    
Before closing, we 
{ emphasize again the importance of the IVP approach for examining the wave dynamics.
The physical relevance of a mode found in some EVP analysis needs to be established
    with an IVP examination before it is employed for seismology.
Likewise, the detailed temporal behavior of an observed oscillatory signal actually
    carries some rich information on both the wave host and the initial perturbation.
Furthermore, we remark}
    that our numerical findings can be summarized in 
    a simpler way, provided that one is interested in 
    the solutions to IVP~\ref{ivp_mu_open} rather than any conceptual
    difficulty associated with the domain being open 
    for some relevant EVPs. 
First of all, the mathematical properties of EVP~\ref{evp_mu_closed} 
    ensure that the associated eigenmodes
    can be used to { solve} IVP~\ref{ivp_mu_open}.
That EVP~\ref{evp_mu_closed} is defined on a closed domain merely means that
    the ``Modal Closed'' solution (Equation~\eqref{eq_modal_formalSol})
    holds only in some appropriate timeframe.
For our chosen parameters,      
    the key difference in the modal behavior between
    { $\mu\ge 2$ and $\mu<2$} is that
    only oscillatory eigenmodes are allowed in the former,
    whereas evanescent eigenmodes eventually show up for sufficiently
    large domains in the latter. 
The { system evolution, however, necessarily shows no qualitative
    difference for different $\mu$} in those timeframes for which
    the required domain sizes of EVP~\ref{evp_mu_closed} are not large enough
    to accommodate evanescent eigenmodes. 
Only oscillatory eigenmodes are therefore involved in
    Equation~\eqref{eq_modal_formalSol}.
The signal at a given location then weakens with time,
    for the monochromatic components will interfere to cancel out their contributions. 
On this last aspect, we note that a close analogy can be drawn for 
    kink waves resonantly absorbed in the \Alf\ continuum
    (\citealt{2015ApJ...803...43S}; see also
    \citealt{1991JPlPh..45..453C} and \citealt{2007PhPl...14e2101A}).

\acknowledgments
{ We thank the referee for thorough and constructive comments.}
This research was supported by the 
    National Natural Science Foundation of China
    (41974200, 41904150, and 11761141002).
We gratefully acknowledge ISSI-BJ for supporting the international team
    ``Magnetohydrodynamic wavetrains as a tool for probing the solar corona''.

\bibliographystyle{aasjournal}
\bibliography{seis_generic}

%

\clearpage

\clearpage
\clearpage

\clearpage

\clearpage

\clearpage

\clearpage

\clearpage

\end{document}